

       \def\A{\cal A}
       \def\AG{{\cal A}/{\cal G}}
       \def\AGbar{\overline{\AG}}
       \def\HA{{\cal HA}}
       \def\HG{\cal HG}
       
       \def\S{\Sigma}
       \def\C{$C^\star$-algebra}
       \def\HA{\cal HA}
       \def\Ab{\bar{A}}
       \def\a{\alpha}
       \def\at{\tilde{\a}}
       \def\b{\beta}
       
       \def\L{{\cal L}_{x_o}}
       \def\SU{[SU(2)]^n/Ad}


       \def\bra#1{\langle\, #1\, |}
       \def\comp{{\rm C}\llap{\vrule height7.1pt width1pt depth-.4pt
        \phantom t}}
       \def\d{\partial}
       \def\half{{\textstyle{1\over2}}}
       \def\implies{\Rightarrow}
       \def\IP#1#2{\langle\, #1\, |\, #2\, \rangle}
       
       \def\lint{\int\nolimits}

       \def\ovr{\overline}
       \def\real{{\rm I\!R}}
       \def\to{\rightarrow}
       \def\Tr{\mathop{\rm Tr}}
       \def\tw{\widetilde}
       \def\Ez{{}^o\!E^a_i}
       \def\Az{{}^o\!A_a^i}

\magnification=1200

\rightline{gr-qc/9403038, CGPG-94/1-1}
\bigskip
{\centerline{\bf Overview and Outlook}}
\medskip
\centerline{Abhay Ashtekar}
\centerline{\it Center for Gravitational Physics and Geometry}
\centerline{\it Pennsylvania State University, University Park, PA 16802-6300}
\medskip

{\narrower\narrower\smallskip\noindent {\sl When you do some
mountaineering...you sometimes...want to climb to some peak but
there is fog everywhere...you have your map or some other indication
where you probably have to go and still you are completely lost in the
fog. Then...all of a sudden you see quite vaguely in the fog, just
a few minute things from which you say, ``Oh, this is the rock I
want.'' In the very moment that you have seen that, then the whole
picture changes completely, because although you still don't know
whether you will make the rock, nevertheless for a moment you say,
``... Now I know where I am; I have to go closer to that and then I
will certainly find the way to go ...'' \smallskip}}
\hskip7cm (Werner Heisenberg, 1963)

{\bf 1. Introduction}

The problem of quantum gravity is an old one and over the course of
time several distinct lines of thought have evolved%
\footnote{.}{To appear in the proceedings of the workshop on the
Canonical Formalism in Classical and Quantum General Relativity,
held at Bad-Honef, Germany in September 1993; H. Friedrich (ed),
Springer-Verlag.}
However, for several decades, there was very little communication
between the two main communities in this area: particle physicists and
gravitation theorists. Indeed, there was a lack of agreement on even
what the key problems are. By and large, particle physics approaches
focused on perturbative techniques.  The space-time metric was split
into two parts: $g_{\mu\nu}=\eta_{\mu\nu} + G h_{\mu\nu}$,
$\eta_{\mu\nu}$ being regarded as a flat kinematic piece, $h_{\mu\nu}$
being assigned the role of the dynamical variable and Newton's
constant $G$ playing the role of the coupling constant. The field
$h_{\mu\nu}$ was then quantized on the $\eta_{\mu\nu}$-background and
perturbative techniques that had been so successful in quantum
electrodynamics were applied to the Einstein-Hilbert action. The key
problems then were those of handling the infinities. The gravity
community, on the other hand, felt that a central lesson of general
relativity is that the space-time metric plays a dual role: it is
important that one and the same mathematical object determine geometry
{\it and} encode the physical gravitational field. From this
perspective, an ad-hoc split of the metric goes against the very
spirit of the theory and must be avoided. If one does not carry out
the split, however, a theory of quantum gravity would be
simultaneously a theory of quantum geometry and the notion of quantum
geometry raises a variety of conceptual difficulties. If there is no
background space-time geometry --but only a probability amplitude for
various possibilities-- how does one do physics? What does causality
mean?  What is time? What does dynamics mean? Gravity theorists
focused on such conceptual issues.  To simplify mathematics, they
often truncated the theory by imposing various symmetry conditions and
thus avoided the field theoretic difficulties.  Technically, the
emphasis was on geometry rather than functional analysis. It is not
that each community was completely unaware of the work of the other
(although, by and large, neither had fully absorbed what the other
side was saying). Rather, each side had its list of central problems
and believed that once these issues were resolved, the remaining ones
could be handled without much difficulty.  To high energy theorists,
the conceptual problems of relativists were perhaps analogous to the
issues in foundations of quantum mechanics which they considered to be
``unimportant for real physical predictions.'' To relativists, the
field theoretic difficulties of high energy physicists were
technicalities which could be sorted out after the conceptual issues
had been resolved. Even on the rare occasions when they got together,
the two sides seemed to ``talk past each other.''

Over the last decade, however, there has been a certain rapprochement
of ideas on quantum gravity. Each side has become increasingly aware
of the difficulties that were emphasized by the other. There are some
agreements and increasing clarity. In the poetic imagery of
Heisenberg, we have seen faint outlines of some of the rocks we want
to climb.
\smallskip

Perhaps the most important among these developments are the following:
\item{1.} Recognition that non-perturbative methods are essential.
There is a growing sentiment that, at a fundamental level, the theory
should not involve fluctuations around a classical geometry. Indeed,
there should be no background geometry or any background fields for
that matter.  The theory should be diffeomorphism invariant.
\item{2.} Acceptance that conceptual issues such as the problem of time
will have to be handled satisfactorily. The unease with the standard
measurement theory of quantum mechanics is brought to the forefront by
the absence of a background, classical space-time. The interface of
classical and quantum domains and the decoherence processes that make
the world seem classical are receiving greater attention.
\item{3.} Agreement that the field theoretic divergences have to be
faced squarely. There is growing awareness that although the
mini-superspaces that feature in quantum cosmology are obviously
interesting, from the perspective of full quantum gravity, they are
essentially toy models. One has to learn to deal with the infinite
number of degrees of freedom ``honestly.'' Time has come to give proper
mathematical meaning to the formal equations such as the Wheeler-DeWitt
equation. New ideas and mathematical techniques are needed since the
issue of regulating quantum operators in absence of a background geometry
has rarely been faced.
\item{4.} Reconciliation with the possibility that the ``quantum
geometry'' governing the Planck scale may not at all resemble the
Riemannian picture. It is likely that differential geometry, in the
standard sense of the term, will itself be inadequate to capture
physics. Some discrete structures are likely to emerge and new
mathematical tools will be needed to handle them. \smallskip
\noindent
Thus, the goals of the two communities have moved closer.

These recognitions do not imply, however, that there is a general
consensus on {\it how} all these problems are to be resolved.  Thus,
there are again many approaches. But I strongly support belief that
this diversity in the lines of attack is very healthy.  In a problem
like quantum gravity, where directly relevant experimental data is
scarce, it would be a grave error if everyone followed the same path.
Indeed, the most promising way of enhancing our chances at success is
to increase the amount of variety. What is striking is that, in spite
of the diversity of methods, some of the results are often qualitatively
similar. For example, in the last few years, the fourth point I
mentioned above has come up again and again in quite different
contexts.  The idea that the continuum picture itself is likely to
break down is not new. For the first time, however, it is arising as a
{\it concrete result} of calculations within well-defined, coherent
schemes which were not explicitly constructed to obtain such a result.
It arose from certain computer simulations of 4-dimensional Euclidean
gravity, (see e.g.  Agishtein \& Migdal (1992) and Br\"ugmann \&
Marinari (1993)), from string theory (see e.g. Gross \& Mende (1988),
Amati et al (1990), Aspinwall (1993)), and, from canonical
quantization of 4-dimensional general relativity (Ashtekar et al
(1992)). The detailed pictures of the micro-structure of space-time
that arise in these approaches are quite different at least at first
sight.  Nonetheless, there {\it are} certain similarities in the
results; most of them are obtained by using genuinely non-perturbative
techniques. The overall situation makes me believe that we have, so to
say, entered a new era in the field of quantum gravity. Through the
dense fog, we have caught a fleeting glimpse of the rocks that, we
think, will bring us to the distant peak and, in the spirit of
Heisenberg, various groups are charting their paths even though they
still do not know if they will make it.

In this article, I will attempt to provide a broad overview of the
canonical approach based on connections and loops and suggest some
directions for future work.  I should emphasize that this is not a
systematic survey. I will not even attempt to cover all areas in which
significant developments have occurred. Rather, I will concentrate on
issues which are related to the topics covered by other speakers but
which were not discussed in their talks in any detail. I should also
emphasize that most of what I will say are personal perspectives which
are not necessarily shared by others in the field.  (For comprehensive
reviews of the canonical approach based on connections and loops see
other articles in this volume and, e.g., Ashtekar (1991,1992) and
Pullin (1993).)

I begin in section 2 with an assessment of the canonical approach and,
using 2+1-dimensional general relativity, illustrate the vision that
underlies the program. In section 3, I will summarize some recent
mathematical developments which were mentioned by Bernd Br\"ugmann
(1994) and Renate Loll (1994) in their talks.  These results have made
it possible to define the loop transform rigorously for non-Abelian
connections with an infinite number of degrees of freedom and they
provide a basis for regulating and solving the quantum constraints.
Section 4 illustrates how low energy (i.e. laboratory scale) physics
can arise from non-perturbative quantum gravity and section 5 presents
some promising directions for future work.

\bigskip\bigskip
\goodbreak
{\bf 2. The canonical approach}

This workshop as a whole was devoted to canonical methods in general
relativity, with a particular emphasis on quantization. Therefore, in
the first part of this section, I will present an assessment of the
strengths and weaknesses of these methods in general terms, without
restricting myself to the choice of a specific type of dynamical
variables or to the details of how the canonical program is carried
out. In the second part, I will present my own view of the general
framework one can hope to arrive at in quantum gravity through the use
of canonical methods.

{\sl 2.1 An assessment}

At the mathematical as well as conceptual level, the quantum theory we
best understand is the non-relativistic quantum mechanics of point
particles. In this case, the canonical approach provides the ``royal
road'' to quantization. One can even formulate the fundamental
kinematic problem as that of obtaining the appropriate representation
of the basic canonical commutation relations. The problem of dynamics
is then that of defining the Hamiltonian operator and of fully
understanding its action on states. Note that the path integral method
is {\it not} a substitute for canonical quantization. It is rather an
alternative --and, in the case of scattering problems, a powerful--
method of tackling the issue of quantum dynamics. What it provides is
the transition amplitudes $\IP{{\vec x'}, t'}{{\vec x}, t}$ for the
particle to go from the position ${\vec x}$ at time $t$ to the
position ${\vec x}'$ at time $t'$. The full quantum mechanical
results, however, involve calculating the amplitude for a transition
from one {\it physically realizable} quantum state to another ---say,
from a state $\Psi_0({\vec x})$ with angular momentum $0$ to a state
$\Psi_m({\vec x})$ with angular momentum $m$. And, to obtain these, we
need to know the Hilbert space of states, the action of physically
interesting operators and the eigenfunctions and eigenvalues of these
operators. These have to be supplied externally ---typically through
canonical quantization. It is for this reason that most courses on
non-relativistic quantum mechanics focus on the canonical method.

Coming to field theory, the canonical approach is again the oldest
coherent method to tackle the problem of quantization. It is deeply
rooted in the Heisenberg uncertainty relation and hence in the
fundamental {\it physical} principles of quantum mechanics. Why is it
then that no modern course in field theory develops the subject within
the canonical method? The main reasons, I believe, are the following.
First, the approach lacks {\it manifest} covariance. Second, it is not
well-suited to the use of Feynman diagrams and other perturbative
techniques. The first drawback arises from obvious reasons: The basic
canonically conjugate fields satisfy equal time commutation relations
and are therefore defined on a Cauchy slice, whence the 4-dimensional
covariance is broken at the outset. The origin of the second drawback
lies in the fact that the quantum states in the canonical approach
arise as suitable functionals of fields and this representation is
quite inconvenient if one wants to work with particle states. In
perturbative field theory, on the other hand, the emphasis is on
particles --real and virtual-- and the powerful machinery of Feynman
diagrams is tuned to the particle picture. Furthermore, each diagram
represents a {\it space-time} process. The exchange of virtual
particles, particles moving forward in time and the anti-particles
moving backwards, real particles scattering off each other --each of
these processes involves the passage of time and is a representation
of a 4-dimensional, space-time integral. It is in principle possible
but in practice cumbersome to capture all this in the canonical
framework. It is hard to prove renormalizability.  It is {\it much}
easier to forego the descriptions in terms of functionals of
3-dimensional fields and consider instead the Fock spaces of particle
states.  Indeed, this is hardly surprising. It is always the case that
calculations simplify when one tailors the framework to the dynamics
in question. In the perturbative treatment of scattering theory, then,
simplifications should occur when we use eigenstates of the
asymptotic, free Hamiltonian. And these are precisely the particle
states in the Fock space.

One might adopt the viewpoint that from the standpoint of what is true
``in principle,'' both the drawbacks stem from aesthetic
considerations.  Indeed, although the procedure is not manifestly
covariant, it is nonetheless true that, in the final picture, the
quantum field theory one obtains, say in the case of a free Maxwell
field, is completely equivalent to the more familiar one featuring the
Fock space of photons. In the canonical approach, the states are
represented by (gauge invariant) functionals $\Psi(A)$ of the vector
potential $A_a({\vec x})$ on a 3-dimensional plane (or, more
generally, a Cauchy surface) in Minkowski space. This description
breaks the manifest covariance.  However, the Poincar\'e group {\it
is} unitarily implemented. The vacuum state {\it is} Poincar\'e
invariant. The spectrum of the (free) 4-momentum operator {\it is}
causal and future-directed, and so on. Similarly, as we noted already,
using the equivalence with the Fock representation, one can translate,
step by step, any perturbative calculation in, say, quantum
electrodynamics to the canonical language. This is all true. But the
price one would pay would be substantial. It would be analogous to
--and {\it much} more complicated than-- insisting on using Cartesian
coordinates in a problem where the Hamiltonian has manifest spherical
symmetry.  Not only would it make calculations enormously more difficult
but one would also lose all sorts of {\it physical} insights. This is
not to say that canonical quantization has no place in field theory.
It does.  Most text books begin the discussion with canonical
quantization perhaps because the underlying basic structure is
clearest in this framework.  Similarly, at the odd place when one is
not sure of the measure in the path integral, one returns to the
canonical picture and ``derives the correct measure'' starting with
the Liouville form on the phase space.  However, for the actual
calculations of scattering cross sections or probing general
properties such as renormalizability, the language of particles and
the 4-dimensional pictures are obviously better-suited.

What about general relativity? The canonical approach has the great
advantage that it does not require us to introduce a background
metric. It is well suited for a non-perturbative treatment. The final
Hamiltonian framework does have a number of features not encountered
in Minkow\-ski\-an field theories. However, in view of the profound
conceptual differences between these theories and general relativity,
the emergence of such features is but to be expected. In particular,
much of the dynamical information of the theory is now contained in
constraints. Indeed, as discussed by Robert Beig (1994), in the
spatially compact case, the Hamiltonian vanishes identically on the
constraint surface, i.e., on the physical states of the classical
theory. In quantum theory then, to begin with, there is no
Hamiltonian, no time and no evolution. One only has physical states
--the solutions to quantum constraints. Yet, it may be possible to
``extract'' dynamics from these solutions by identifying a suitable
physical variable as an internal clock. Indeed, this has already been
done in 2+1-dimensional gravity (see e.g. Carlip (1993)). Thus, at
least in principle, the approach provides us with a well-defined,
precise strategy that is sufficiently sophisticated to extract physical
information at the quantum level in spite of the absence of a background
geometry. This is its greatest strength.

How does the approach fare with respect to the two key difficulties it
faces in Minkow\-ski\-an quantum field theories?  In quantum gravity,
the scattering cross sections are not our {\it prime} concerns. Since
there is no background space-time metric, notions such as particle
states --and gravitons in particular-- which are tied to the
Poincar\'e group would presumably be only approximate concepts. The
fact that quantum general relativity (and various modifications
thereof) fails to provide us with a consistent, local quantum field
theory perturbatively also indicates that the fundamental theory
should not be formulated in terms of these particle states. The whole
imagery of processes mediated by particles moving forward and backward
in time is probably inappropriate. So, the second reason for
abandoning the canonical method now loses its force.  Indeed, the key
question now is if the quantum dynamics of the gravitational field can
be made simple in an appropriate representation within the canonical
scheme. And, as we have seen in this workshop, there are indeed strong
indications that the loop representation is well-suited for this
purpose (see e.g.  Br\"ugmann (1994)).

The lack of manifest covariance of the canonical scheme, however, is
still with us.  And indeed it is now a {\it much more} serious issue.
In Minkowskian field theories the issue was an aesthetic one. While
the canonical procedure violates {\it manifest} covariance, it {\it
is} covariant; the Poincar\'e group {\it is} unitarily implemented. We
can, if we wish, describe dynamics in the space-time picture.  In
quantum gravity this is no longer the case. In classical general
relativity, it is the space-time geometry that is the dynamical
variable. A space-time represents a possible history --analogous to a
trajectory in particle mechanics. Just as in non-relativistic quantum
mechanics the particle trajectories have no basic role to play in the
final quantum description, one would expect that in full canonical
quantum gravity, space-times would have no distinguished place. There
will presumably exist some special (``semi-classical'') states which
can be approximated by 4-dimensional space-times. Given any such
state, one would be able to speak of approximate 4-dimensional
covariance in an effective theory which ignores large quantum
flucutations away from that state. At a fundamental level, however,
there would not be a 4-dimensional geometric entity to replace the
classical space-time.  And if there is no such entity, how can one
even speak of 4-dimensional covariance? It does seem that there is a
branching of ways here.  Furthermore, it is the qualitative features
of the canonical approach that force this branching; it is not tied to
the use of specific variables.

To preserve the space-time covariance, it is tempting to choose the
path integral approach. Even though in the finished picture there is
no preferred space-time, one does have an underlying 4-manifold and
the basic objects that one calculates with are 4-metrics. And the
diffeomorphism group does act on the space of these metrics. Since the
Einstein-Hilbert action is diffeomorphism invariant, one would expect
to get diffeomorphism invariant answers for (the correctly formulated)
physical questions. Thus, the strategy seems attractive. There are,
however, three problems.

First, since path integrals in full quantum gravity are only formal
expressions, everything one does {\it from the very beginning}
involves formal manipulations; the level of mathematical precision
leaves much to be desired. In practice, one generally proceeds by
making some ``approximations'' which mimic the successful strategies
in other field theories. However, it is not at all obvious that these
strategies can be taken over to gravity in a meaningful fashion.
Indeed, in Minkowskian field theories such as quantum electrodynamics,
the only way we know how to make sense of path integrals is through
perturbation theory. In fact, the power of the method lies in the fact
that it captures the enormous information in the perturbation
expansions very succinctly. In a theory that does not exist
perturbatively, we are at a loss. It would be surprising indeed if
all of the approximation methods that are useful in the perturbative
context will continue to be useful in quantum gravity. A much more
sophisticated approach is needed.  One might imagine using the
techniques developed in rigorous quantum field theory. However, in
constructive quantum field theories, path integrals are defined in the
Euclidean regime and the physical, Lorentzian Green's functions are
obtained through a wick rotation. We all know that, in gravity, this
simple route is not available.

The second problem is that while the method is manifestly covariant in
the sense of classical general relativity, it lacks {\it quantum
covariance} even in the case of particle dynamics. Let me explain this
point in some detail since many of the readers may be unfamiliar with
the issue. In the sum over histories approach, one fixes, at the
outset, a preferred configuration space ---all histories are to be
trajectories in this space.  In quantum mechanics, this corresponds to
fixing a representation (such as the position or the momentum) of the
observable algebra. Now, one of the technically powerful and, I feel,
conceptually deep features of quantum physics is Dirac's transformation
theory%
\footnote{${}^\dagger$}{about which Dirac said in 1977: ``I think that
is the piece of work which has most pleased me of all the works that I
have done in my life...The transformation theory became my darling.''
(Pais (1987))}
which establishes the covariance of the {\it quantum} theory under the
change of representation. This covariance is manifest in the canonical
approach. In the path integral approach, not only it is not manifest
but it is often cumbersome to incorporate.  Consequently, just as it
is often hard to phrase ``space-time questions'' in the canonical
approach, it is hard to phrase questions which involve different
representations in the path integral approach (unless of course one
passes through the canonical approach and makes use of the Dirac
transformation theory). Consider for example an harmonic oscillator.
If one wants to ask for the probability for the particle which passes
through the interval $\Delta_0$ of the position space at time $t_0$
and the interval $\Delta_1$ at time $t_1$ to end up in an interval
$\Delta_2$ at time $t_2$, we immediately know we should integrate over
the set of paths which pass through the given intervals at the given
times.  Now suppose we change the question somewhat by replacing the
condition at the intermediate time $t_1$ and ask instead that, at that
time, the particle have an energy $\hbar\omega(n+\textstyle {1\over
2})$ for a fixed $n$. Now, it is no longer obvious what paths to
consider. More generally, paths or histories are trajectories in {\it
some} configuration space (i.e. a Lagrangian sub-manifold of the phase
space) and path integrals can easily cope with questions that refer to
that fixed sub-space. In quantum theory, by contrast, the domain space
of wave functions may have nothing to do with the classical phase
space. For example, in the case of the harmonic oscillator, we could
use the energy representation in which the quantum states are
functions $\Psi(n)=\IP{n}{\Psi}$ of an integer which (do not
constitute a Lagrangian sub-manifold of the phase space and therefore)
can not serve as a classical configuration space. Path integrals are
not well-adapted to such representations. But it may well be that it
is precisely such representations that are best suited to incorporate
quantum dynamics. The loop representation, in particular, falls in
this category.

The last problem is the one I already mentioned in the beginning of
this sub-section. Path integrals provide transition amplitudes but not
the kinematical structure; it has to be supplied from outside. In a
real sense, therefore, the path integral and the canonical approaches
generally complement each other. In quantum gravity, it is possible
that ultimately the two will be used together so that the framework
has both classical and quantum covariance.  The dynamics coded in the
quantum constraints in the canonical approach could perhaps be
re-interpreted in terms of mathematically well-defined path integrals
to demonstrate that there is a well-defined sense in which the theory
enjoys space-time covariance. Such an interpretation would be
especially helpful in the analysis of whether topology changes occur
in quantum gravity and, if so, whether they have physically
significant ramifications in the low energy regime. The full quantum
covariance could be established by switching to the canonical
framework.

It may also turn out, however, that in the gravitational case,
canonical quantization can not be reconciled with path integrals in
the full quantum theory. As the discussion in the next sub-section
indicates, even the ``spatial'' 3-manifold could be a secondary
construct. At the fundamental level, there may only be discrete
structures and combinatorial operations.  In such a scenario, the
continuum picture and objects such as manifolds may arise as useful
mathematical constructs only in semi-classical physics. And the theory
would have the desired covariance only in these regimes. It may be that
the predictions of the two methods agree in such regimes but in the fully
Planckian domain, the two theories are quite different.

To summarize then, the canonical methods are well adapted to
non-perturbative treatments of quantum gravity. However, there appears
to be a fundamental tension between the quantum gravity scenarios that
are natural to canonical approaches and space-time pictures that we
are so accustomed to in the classical regime. Whether this tension is
real or only apparent is not yet clear. Technical progress,
particularly in the path integral approach, would be of great help to
settle this issue.

\bigskip\goodbreak
{\sl 2.2 An underlying vision}

In this workshop, Thomas Thiemann discussed 2+1-dimensional gravity in
some detail and Hermann Nicolai presented an extension (due to de Wit,
Matschull and himself (1993)) of those results to give an elegant
clarification of the issue of the ``size'' of the space of physical
states in supergravity. I would now like to use 2+1 gravity for a
different purpose: to illustrate my own expectations of
non-perturbative quantum gravity in 3+1 dimensions.  (For details on
2+1-dimensional gravity, in addition to these proceedings, see, e.g.,
Carlip (1990, 1993), and Ashtekar (1991), chapter 17.)

Let me begin with a brief historical detour.  Conceptually --and, in
certain respects, also technically-- 2+1-dimensional general
relativity is very similar to the 3+1-dimensional theory. There is no
background structure; the theory is diffeomorphism invariant. One and
the same object --the space-time metric $g_{\mu\nu}$-- determines the
geometry {\it and} encodes the gravitational field. In the canonical
description, the Hamiltonian again vanishes in the spatially compact
case and the dynamics is driven by first class constraints.
Furthermore, in geometrodynamics, the general structure of the
constraint algebra is the same as in the 3+1 theory (see, e.g., Beig
1994); it is not a true Lie algebra.  Consequently, for a number of
years, it was believed that a non-perturbative, canonical quantization
of this theory is as difficult as that of the 3+1-dimensional theory.
In particular, not a single solution to the Wheeler-DeWitt equation
was known.  In the perturbative treatment, one finds that Newton's
constant is again dimensionful (with dimension [L], thus a {\it
positive} power of length in the $\hbar=c=1$ units) and simple power
counting arguments suggested that the theory is not renormalizable.
Therefore, there appeared several papers in the literature saying that
2+1-dimensional general relativity is as intractable as the
3+1-dimensional theory.

This would have been somewhat puzzling since in 2+1-dimensions there
are no local degrees of freedom; all solutions to the Einstein field
equations are flat. Indeed, the conclusions of these papers turned out
to be wrong. There are no divergences. Using connections as basic
variables rather than metrics, the quantum theory can in fact be
solved exactly in the spatially compact case for any genus. (Achucarro
\& Townsend (1986), Witten (1988). See also Ashtekar et al (1989) and
Nelson \& Regge (1989)). Since then, the theory has also been solved
in the metric representation in the case when the spatial topology is
that of a 2-torus (Moncrief 1989, Hosoya \& Nakao (1990)) and work is
in progress on the higher genus case. One of the lessons that one can
draw from this work is that connections and holonomies are better
suited to the mathematics of the quantum theory than metrics and light
cones. It is {\it not} that the theory can not be solved in the metric
picture; as I just remarked, this has already been achieved in the
2-torus topology. Rather, connections seem to be better adapted to
deal with the quantum constraints and to write down Dirac observables.
Variables which are most useful to the macro-physics are not the ones
which make the micro-physics most transparent.

There are several equivalent ways of quantizing 2+1 gravity on a torus
(see, e.g., Carlip 1993). The variety of representations that
result provide us with a rich example of the Dirac transformation
theory in action. Here, I will consider three and argue that each
clarifies and emphasizes a different aspect of quantum gravity.
At first sight, it appears that each representation provides its own,
distinct picture of reality. And yet, thanks to the Dirac
transformation theory, they are all equivalent.

First, there is the loop representation. Here, quantum states are
functions of homotopy classes of loops on the spatial 2-manifold; the
domain space of quantum states is thus {\it discrete}. In the case of
a torus, each homotopy class is represented by a pair of integers
$(n_1, n_2)$ which tell us how many times the loop winds around the
two generators of the homotopy group. Hence, in this case, the quantum
states $\Psi(n_1, n_2)$ are functions just of two integers. The basic
observables of the theory, $T^0(n_1, n_2)$ and $T^1(n_1,n_2)$, are
also labelled by two integers and their action on states just shifts
the arguments of the wave functions. Thus, the whole mathematical
structure is {\it combinatorial}.  There is no space, no time, no
continuum. Hence there is no diffeomorphism group to implement and no
issue of space-time covariance to worry about. I would like to regard
this as the ``fundamental description.''  Now, we learnt from Thomas
Thiemann's talk (1994) that this description is completely equivalent
to that in terms of connections on a 2-torus%
\footnote{${}^\dagger$}{The two are related by the loop transform. A
certain amount of caution is needed in the definition of the loop
transform however because the space on which one integrates --the
moduli space of flat connections-- has non-compact connected components
on which the Wilson loop functions are unbounded (Marolf 1993, Ashtekar
\& Loll (1993)). This important technical point was overlooked in the
earlier treatment (Ashtekar et al 1989).}.
Put differently, suppose we knew nothing about general relativity but
were provided with just the combinatorial description given above.
Staring at this description, a clever young student could have
realized that the description would become ``nice and geometric'' if
one were to introduce a ``fictitious, mathematical'' 2-manifold,
namely a 2-torus, and regard the arguments $(n_1, n_2)$ of quantum
states as labels for the homotopy classes of closed loop on this
torus. Excitedly, she tells us that, once this is done, one can
equally consider the quantum states of two real numbers $(a_1, a_2)$
which can be regarded as labels for gauge equivalent flat connections
on the torus.. After working in this picture for a while, we realize
that the theory can be regarded as quantization of an interesting
classical theory which features connections on a 3-dimensional
manifold, obtained by taking the product of our fictitious torus with
a fictitious real line. To our astonishment, we then notice that the
theory has the peculiar property of being diffeomorphism invariant.
Furthermore, we notice that it can be thought of as a theory of
metrics of signature $- + +$ and the connection which we discovered
earlier is the spin connection of this metric. We can then go on and
study the light cones and causal properties of this classical theory
and be surprised that all this rich and unexpected structure arises in
the classical limit. We discover that, in this classical
approximation, the 3-manifold can be thought of as ``space-time'' in
which things live.  Intrigued by this notion of time and dynamics, we
return to the quantum theory and realize that with a little bit of
work, we can construct a third representation in which quantum states
are functions of {\it three} variables subject to a differential
equation which tells us how the state changes or ``evolves'' as we
increase the third variable. We realize that variable could be
interpreted as an ``internal clock.''

Thus, we have three different quantum pictures.  The ``fundamental''
description is combinatorial. The second, in terms of connections, is
a ``timeless'' description. We can interpret the 2-manifold as
``space'' and see that the spatial diffeomorphisms are unitarily
implemented. But there is no time; we have a ``frozen formalism'' a la
Bergmann and Komar. Finally, in the third description, we reintroduce
time through an internal clock. Now the description resembles that of
a parametrized particle.  We are on familiar ground.

My hope is that we will arrive at a similar description in the 3+1
theory. There will be several equivalent representations, each
illuminating its favorite facets of quantum gravity. Thanks to the
Dirac transformation theory, the numerous expectations we have of the
theory --which often seem even to contradict each other-- will in fact
be compatible. There is already some evidence in favor of this
scenario. The loop and connection representations are qualitatively
similar to those in the 2+1-dimensional case. (In fact, the emphasis
on connections in the 2+1-dimensional theory was motivated, in the
first place, by the connection dynamics formulation of the 3+1
theory.)  In the loop representation, the diffeomorphism constraint
leads us to knot classes and hence to combinatorial operations. The
connection representation will lead us directly to a timeless, frozen
formalism.  And as we will see in section 4, in the weak field
truncation, we can identify one of the components of the connection as
an internal clock and recover quantum dynamics from the Hamiltonian
constraint. Thus, we do have a number of pointers. Of course, whether
this scenario is borne out in detail remains to be seen. In my view,
the weakest point may be the last one; we may be able to isolate
internal clocks only under suitable approximations.

\bigskip\bigskip
\goodbreak
{\bf 3. Recent mathematical developments}

It is obvious from the articles by Br\"ugmann, Loll and Thiemann that
the loop representation plays an important role in this approach to
quantum gravity. It is therefore important to put this representation
and results obtained therefrom on a sound mathematical footing. It
is worth emphasizing that this is not just an abstract desire for
rigor.  In a real sense, in non-perturbative quantum gravity, we are
entering a completely uncharted territory and the mathematical
techniques required are quite new. In most of the work to date, one
has proceeded by making convenient choices --e.g., of the regularity
conditions on loop states, of factor-ordering and of limiting
procedures used to regulate operators-- as the need arose. These
choices do seem ``reasonable'' and there {\it is} an overall coherence
to the entire scheme. However, it is not clear that there aren't other
choices that are equally plausible and that the results will not
change if these choices were made.  There are no uniqueness results.
And none will come by unless one elevates the degree of mathematical
precision. In a sense, this is analogous to the situation in
interacting quantum field theory in Minkowski space, where there is
again no a priori justification for using the Fock representation and
the associated regularity conditions on states and operators. (In
fact, in view of Haag's theorem (see, e.g., Streater \& Wightman
(1964)), there are good mathematical reasons for {\it not} using it!).
However, in that case, over the years one has acquired a great deal of
theoretical experience and, more importantly, there is enormous
experimental data which supports the strategy as a working hypothesis.
In non-perturbative gravity, on the other hand, one lacks both and the
issue of control over the assumptions and uniqueness results
become much more important.

Fortunately, over the last two years several mathematical developments
have occurred which have the potential of improving the situation
dramatically. The finished and published results deal only with real
$SU(n)$ or $U(n)$ connections. However, work is in progress to extend
these results to the complex-valued connections that one encounters in
gravity and it appears that the key results which enable one to define
the loop transform and the loop representation will go over to that
case as well%
\footnote{${}^\dagger$}{The idea is to let the gauge group be $SU(2)$
--as is needed for the triads $E^a_i$ to be real-- but to allow
complex-valued connections $A_a^i$ (which take values in the
Lie-algebra of $\comp SU(2)$, complexified $SU(2)$. Thus, we are using
the original Hamiltonian framework in which $A_a^i$ is regarded as a
complex-valued coordinate on the real phase space of general
relativity. One then uses appropriate extensions of the techniques
outlined below to develop calculus on the resulting $\AG$. In
particular, one can give a precise meaning to the notion of
holomorphic functionals used in the connection representation. We
expect that the appropriate parts of the integration theory will also
go through and enable us to define the loop transform rigorously,
although here the work is still incomplete.  If all our expectations
are borne out, we would also have incorporated the reality conditions
at the kinematical level, i.e.  prior to the imposition of
constraints.}.
For brevity and precision, in this section, I will restrict myself to
$SU(2)$ connections and report on the results that have been already
obtained.

To define the loop transform, we need an integration theory on the
space $\AG$ of connections modulo gauge transformations. In the
2+1-dimensional case, this problem is easy to solve because the
(appropriate components of the) moduli space of {\it flat} $SU(1,1)$
connections can be naturally given the structure of a finite
dimensional symplectic manifold (see e.g. Ashtekar (1991), chapter 17);
one can simply use the Liouville volume element to evaluate integrals.
Thus, although the domain space of quantum states is non-linear, the
integration theory is simple because the space is {\it finite}
dimensional.  In 3+1 space-time dimensions, the situation is again
simple for the case of {\it linearized gravity} --the theory of free
gravitons in Minkowski space.  This theory can be cast in the language
of connections (see, e.g., Ashtekar (1991), chapter 11).  Integration
theory is again well-developed; the domain space is now linear and one
can simply use the Gaussian measure as in free field theories in
Minkowski space.  Thus, in this case, in spite of the presence of an
infinite number of degrees of freedom, the integration theory is
straightforward because of the underlying {\it linearity}.  The
problem is significantly more difficult if $\AG$ is {\it both}
non-linear {\it and} infinite dimensional.

Let us now discuss this case. Let us begin by fixing an analytic
3-manifold $\S$ which will of course represent a Cauchy surface in
space-times to be considered. We will consider $SU(2)$ connections on
$\S$.  Since any $SU(2)$ bundle over a 3-manifold is trivial, we can
represent any connection by a Lie-algebra valued 1-form $A_a^i$ on
$\S$, where $a$ is the spatial index and $i$, the internal%
\footnote{${}^\dagger$}
{Results reported in this section for which explicit references are
not provided are all taken from Ashtekar and Isham (1992) and Ashtekar
and Lewandowski (1993a,b). The last two papers and those by Baez
(1993a,b) contain significant generalizations which include allowing
more general gauge groups, allowing the manifold $\S$ to be of
arbitrary dimension and allowing the connections to live in
non-trivial bundles.}.
Denote by $\A$ the space of smooth (say $C^2$) $SU(2)$-connections
equipped with one of the standard (Sobolev) topologies (see, e.g.
Mitter and Viallet (1981)). $\A$ has the structure of an affine space.
However, what is of direct interest to us is the space $\AG$ obtained
by taking the quotient of $\A$ by $(C^3)$ local gauge transformations.
In this quotient construction, the affine structure is lost; $\AG$ is
a genuinely non-linear space with complicated topology.

To define the integration theory, we will adopt an algebraic approach.
To see the main idea, consider integrals $\int fd\mu$ of continuous
functions $f$ on a compact, Hausdorff space $S$ with respect to a
regular (but not necessarily strictly positive) measure $d\mu$. The
space of continuous functions has the structure of an Abelian
$C^\star$-algebra with identity and the map $f\to <f>:= \int f d\mu$
is a functional on the algebra satisfying the following two
properties: i) positivity (which should really be called
non-negativity), i.e. $<\bar{f}f>\ \ \ge 0$ for any element $f$ in the
algebra; and, ii) linearity, i.e. $<f+\lambda g> = <f>+\lambda <g>$ for
all elements $f$ and $g$ of the algebra and complex numbers $\lambda$.
Thus, given a measure, we acquire a positive linear functional on the
given Abelian \C. Now, standard theorems ensure us that {\it every}
regular measure on $S$ arises in this manner. The idea now is to use
this 1-1 correspondence to define measures on the space of connections
modulo gauge transformations.

Let us begin by constructing an appropriate sub-algebra of the Abelian
$C^\star$-algebra of continuous, bounded functions on $\AG$. This is
to be the algebra of configuration variables. Therefore, the obvious
strategy is to use the Wilson loop functionals. Following the
procedure and notation used by other speakers, given any closed loop
$\a$ on the 3-manifold $\S$, let us define the Wilson-loop functional
$T_\a$ on $\AG$:
$$T_\a(A):= \half \Tr\ {\cal P}\ \exp\ G\oint_\a A.dl, \eqno(3.1)$$
where the trace is taken in the fundamental representation of $SU(2)$
and the Newton's constant $G$ appears because, in general relativity,
it is $GA_a^i$ that has the dimensions of a connection; in gauge
theories, of course, this factor would be absent. For technical
reasons, we will have to restrict ourselves to piecewise analytic
loops $\a$. (This is why we needed $\S$ to be analytic. Note that the
loops need not be smooth; they can have kinks and self-intersections but
only at a finite number of points.) As explained in detail by
Br\"ugmann and Loll, due to $SU(2)$ trace identities, the product of
any two Wilson-loop functionals can be expressed as a sum of other
Wilson loop functionals. Therefore, the vector space generated by
finite complex-linear combinations of these functions has the
structure of a $\star$-algebra (where the $\star$-operation is just
complex-conjugation). The functionals $T_\a$ are all bounded (between
$-1$ and $1$). Hence, the sup-norm (over $\AG$) is well-defined and we
can take the completion to obtain a \C. We will call it the {\it
holonomy \C} and denote it by $\HA$. Elements of $\HA$ are to be
thought of as the configuration variables of the theory.

Since $\HA$ is an Abelian $C^\star$-algebra with identity, we can
apply a standard theorem due to Gel'fand and Naimark to conclude that
$\HA$ is isomorphic with the $C^\star$-algebra of all continuous
functions on a compact, Hausdorff space $sp(\HA)$, the spectrum of the
given $C^\star$-algebra $\HA$. The spectrum itself can be recovered
from the algebra $\HA$ directly; its points are homomorphisms from
$\HA$ to the $\star$-algebra of complex numbers. Now, we know that the
elements of $\HA$ suffice to separate points of $\AG$; given any two
elements of $\AG$ there is at least one loop $\a$ such that $T_\a$
takes different values on the two elements.  Therefore it follows
from the Gel'fand-Naimark theorem that $\AG$ is {\it densely embedded
in} $sp(\HA)$.  To emphasize this point, from now on, we will denote
the spectrum by $\AGbar$ and regard it as a completion of $\AG$ (in
the Gel'fand topology). Integration theory will be defined on $\AGbar$
rather than $\AG$. This enlargement is in accordance with the common
occurrence in quantum field theory: while the classical configuration
(or phase) space may contain only smooth fields (typically taken to
belong to be the Schwartz space), the domain space of quantum states
is a completion of this space in an appropriate topology (the space of
distributions). From the remarks I made above, it is clear that
regular measures on $\AGbar$ will correspond to positive linear
functions on the holonomy algebra $\HA$. This is the  general strategy
we will follow.

However, a key difficulty with the use of the Gel'fand theory is that
one generally has relatively little control on the structure of the
spectrum. A reasonable degree of control is essential since (at the
kinematical level) the quantum states in the connection representation
are to be functions $\Psi(\Ab)$ on $\AGbar$. In the present case, we
are fortunate: a simple and complete characterization of the spectrum
is available.  To present it, I first need to introduce a key
definition.  Fix a base point $x_o$ in the 3-manifold $\S$ and regard
two (piecewise analytic) closed loops $\a$ and $\a'$ to be equivalent
if the holonomy of any connection in $\A$, evaluated at $x_o$, around
$\a$ is the same as that around $\a'$. We will call each equivalence
class (a holonomically equivalent loop or) a {\it hoop} and denote the
hoop to which a loop $\a$ belongs by $\at$. For example, $\a$ and
$\a'$ define the same hoop if they differ by a reparametrization or by
a line segment which is immediately re-traced.  (For piecewise
analytic loops and $SU(n)$ connections, one can show that these two
are the most general operations; two loops define the same hoop if and
only if they are related by a combination of reparametrizations and
retracings.)  The set of hoops has, naturally, the structure of a
group. We will call it the {\it hoop group} and denote it by $\HG$. In
terms of this group, we can now present a simple characterization of
the Gel'fand spectrum $\AGbar$:
{\smallskip\narrower{\sl \noindent
Every homomorphism $\hat{H}$ from the hoop group $\HG$ to the gauge
group $SU(2)$ defines an element $\Ab$ of the spectrum $\AGbar$ and
every $\Ab$ in the spectrum defines a homomorphism $\hat{H}$ such that
$\Ab(\at) = \textstyle {1\over 2} \Tr\ \hat{H}(\at)$. This is a 1-1
correspondence modulo the trivial ambiguity that homomorphisms $\hat{H}$
and $g^{-1}\cdot \hat{H}\cdot g$ define the same element $\Ab$ of the
spectrum.}\smallskip}
\noindent Clearly, every regular connection $A$ in $\A$ defines the
desired homomorphism simply through the holonomy operation:
$\hat{H}(\at) := {\cal P}\ \exp G\oint_\a A.dl$, where $\a$ is any
loop in the hoop $\at$. However, there are many homomorphisms which
do not arise from smooth connections. This leads to ``generalized
connections'' --i.e. elements in $\AGbar -\AG$. In particular, there
exist $\Ab$ in $\AGbar$ which have support at a single point and are
thus ``distributional.'' Note that this characterization of the
spectrum $\AGbar$ is {\it completely algebraic}; there is no continuity
assumption on the homomorphisms.  This property makes the
characterization very useful in practice.

As noted above, positive linear functionals on $\HA$ are in 1-1
correspondence with regular measures on (the compact Hausdorff space)
$\AGbar$. It turns out that the positive linear functions, in turn,
are determined completely by certain ``generating functionals''
$\Gamma(\a)$ on the space $\L$ of loops based at $x_o$:
{\smallskip\narrower{\sl \noindent There is a 1-1 correspondence
between positive linear functionals on $\HA$ (and hence regular
measures on $\AGbar$) and functional $\Gamma(\a)$ on $\L$ satisfying:
\hfill\break
\noindent i) $\sum_i a_i T_{\a_i} = 0 \implies \sum_i a_i \Gamma(\a_i)
= 0$; and, \hfill\break
\noindent ii) $\sum_{i,j}\ovr{a}_{i} a_j(\Gamma(\a_i\circ\a_j)+
\Gamma(\a_i \circ\a_j^{-1} )\ge 0$.\hfill\break
\noindent for all loops $\a_i$ and complex numbers $a_i$.}
\medskip}
\noindent The first condition implies that the functional $\Gamma$ is
well-defined on hoops. Hence we could have taken it to be a functional
on $\HG$ from the beginning. Thus, we see that there is a nice
``non-linear duality'' between the spectrum $\AGbar$ and the hoop
group $\HG$: Elements of $\AGbar$ are homomorphisms from $\HG$ to
$SU(2)$ and regular measures on $\AGbar$ correspond to certain
functionals on $\HG$. Finally, if one is interested in measures on
$\AGbar$ which are invariant under the (induced) action of
diffeomorphisms on $\S$, one is led to seek functionals $\Gamma(\a)$
which depend not on the individual loops $\a$ but rather on the
(generalized) knot class to which $\a$ belongs. (The qualification
``generalized'' refers to the fact that here we are allowing the loops
to have kinks, overlaps and self-intersections. Until recently, knot
theorists considered only smoothly embedded loops.) Thus, there is an
interesting --and potentially powerful-- interplay between knot theory
and representations of the holonomy algebra $\HA$ in which the
diffeomorphism group of $\S$ is unitarily implemented.

Finally, we can make the integration theory more explicit as follows.
Consider a subgroup $S_n$ of the hoop group $\HG$ which is generated
by n independent hoops. We can introduce the following equivalence
relation on $\AGbar$: $\Ab\approx \Ab'$ if and only if their action on
all elements of $S_n$ coincides, i.e., if and only if $\Ab(\at) =
g^{-1}\cdot\Ab'(\at)\cdot g$ for all $\at \in S_n$ and some (hoop
independent) $g\in SU(2)$. Thus, intuitively, to be equivalent, $\Ab$
and $\Ab'$ have to agree at points on the set of loops in $S_n$; their
behavior elsewhere on $\S$ does not matter.  {\it It turns out that
the quotient space is isomorphic to} $\SU$. This is the key
result that enables us to make the space of integrable functions and
the integration procedure itself more transparent. It says that the
quotients obtained using two different sub-groups of the hoop group
are isomorphic if and only if the two sub-groups are generated the
same number of independent hoops.  The details of the subgroups do not
matter. There is a certain ``universality'' to the structure of the
quotient.

We want to be able to integrate the generalized Wilson loop
functionals $T_{\at}(\Ab) := \Ab(\at)$ on $\AGbar$.  Using the
quotient construction given in the last paragraph, we will first
present a convenient characterization of the space of all integrable
functions on $\AGbar$.  Let us begin with a definition.  A function
$f$ on $\AGbar$ will be said to be {\it cylindrical} if it is the
pull-back to $\AGbar$ of a smooth function $\tw{f}$ on $\SU$ for some
sub-group $S_n$ of the hoop group. These will be the integrable
functions and they include the generalized Wilson loops. Next, we
define integrals of these functions $f$ on $\AGbar$ explicitly using
the integrals of $\tw{f}$ on $\SU$ , provided of course we equip $\SU$
with suitable measures $d\mu_n$ for each $n$.  We can then first
define a positive linear functional $\Gamma'$ on the space of
cylindrical functions $f$ via:
$$ \Gamma'(f) := \int_{\SU} \tw{f} d\mu_n\ .
\eqno(3.2)$$
For the functional to be well-defined, of course, the family of
measures $d\mu_n$ on $\SU$ must satisfy certain consistency
conditions.  It turns out that these requirements can be met.  The
resulting functionals $\Gamma'(f)$ define regular measures $d\mu'$ on
$\AGbar$ such that $\Gamma'(f) = \int\ f d\mu'$.

A particularly natural choice (and, not surprisingly, the first to be
discovered) is to let $d\mu_n$ be simply induced on $\SU$ by the
Haar-measure on $SU(2)$. With this choice, we have the following
results: {\smallskip\narrower{\sl
\noindent i)The consistency conditions are satisfied; the left side of
(3.2) is well-defined for all cylindrical functions $f$ on $\AGbar$;
\hfill\break
\noindent ii) The generalized holonomies $T_{\at}$ are cylindrical
functionals on $\AGbar$ and $\Gamma(\a): = \Gamma' (T_{\at})$ defined
via (3.2) serves as a generating functional for a faithful, cyclic
representation of the honomony $C^\star$-algebra $\HA$ which ensures
that $d\mu$ is a regular, strictly positive measure on $\AGbar$;
\hfill\break
\noindent iii) The measure $d\mu$ is invariant under the induced action
of the diffeomorphism group on $\S$.}\smallskip}
(The knot invariant defined by $d\mu$ is a genuinely generalized one;
roughly, it counts the number of self-overlaps in any given loop.)

This measure is in some ways analogous to the Gaussian measure on
linear vector spaces. Both can be obtained by a ``cylindrical
construction.'' The Gaussian measure uses the natural metric on
$\real^n$ while the above measure uses the natural (induced) Haar
measure on $\SU$. They are both regular and strictly positive. This
leads us to ask if other properties of the Gaussian measure are
shared.  For instance, we know that the Gaussian measure is
concentrated on distributions; although the smooth fields are dense in
the space of distributions in an appropriate topology, they are
contained in a set whose total measure is zero. Is the situation
similar here? The answer turns out to be affirmative. The classical
configuration space $\AG$ with which we began is dense in the domain
space $\AGbar$ of quantum states in the Gel'fand topology.  However,
$\AG$ is contained in a set whose total measure is zero. The measure
is again concentrated on ``generalized'' connections in $\AGbar$
(Marolf \& Mour\~ao (1993)).  In a certain sense, just as the Gaussian
measures on linear spaces originate in the harmonic oscillator, the
new measure on $\AGbar$ originates in a (generalized) rotor (whose
configuration space is the $SU(2)$ group-manifold).  However, the
measure is, so to say, ``genuinely'' tailored to the underlying
non-linearity. It is {\it not} obtained by ``perturbing'' the Gaussian
measure.

With the measure $d\mu$ at hand, we can consider the Hilbert space
$L^2(\AGbar, d\mu)$ and introduce operators on it. This is {\it not}
the Hilbert space of physical states of quantum gravity since we have
not imposed constraints. It is a fiducial, {\it kinematical} space
which is to enable us to regularize various operators (in particular,
the quantum constraint operators). The configuration operators are
associated with the generalized Wilson loop functionals:
$\hat{T}_{\at}\circ\Psi(\Ab) = \Ab(\at)\Psi(\Ab)$. One can show that
there are bounded, self-adjoint operators on the Hilbert space. There
are also ``momentum operators'' --associated with closed,
2-dimensional ribbons or strips in the 3-manifold $\S$-- which are
gauge invariant and linear in the electric field. One can show that
these are also self-adjoint (but unbounded). Finally, since $d\mu$ is
invariant under the induced action of the diffeomorphism group of
$\S$, this group acts unitarily.

Next, we can now make the Rovelli-Smolin loop transform (Rovelli \&
Smolin (1990), see, also Gambini \& Trias (1986), Br\"ugmann (1994),
Loll (1994)) rigorous.  Since each generalized Wilson loop function
$T_{\at}(\Ab)$ is cylindrical and belongs to the Hilbert space
$L^2(\AGbar, d\mu)$, the loop transform can be defined simply
as a scalar product:
$$\psi({\at}) := \int_{\AGbar} \overline{T_{\at}}({\Ab})
\Psi({\Ab}) \ d\mu\ .\eqno(3.3)$$
As desired, it enables us to pass from functionals $\Psi(\Ab)$ of
generalized connections to functionals $\psi(\at)$ of hoops%
\footnote{${}^\dagger$}{Since the left side is a function of hoops,
we should, strictly, use the terms hoop representation and the hoop
transform.  However, in various calculations, it is often convenient
to lift these functionals from the hoop group $\HG$ to the space of
loops $\L$.  Therefore, as in most of the literature on the subject,
we will not keep a careful distinction between loops and hoops in what
follows.}.
With the rigorously defined transform at hand, we can take over
various operators from the connection to the loop representation. For
the Wilson loop operators $\hat{T}^0_{\at}$ as well as the strip
operators --which are the smeared versions of $\hat{T}^{a}_{\at}(s)$--
we recover the same answers as have been used in all the calculations
in the loop representation. For the operator $\hat{T}^{ab}(s,t)$,
which is used in the regularization of the metric and of the
Hamiltonian constraint operators, work is still in progress.

Finally, let us consider the constraints. Our task here is to
represent the quantum constraints as well-defined operators on the
Hilbert space $L^2(\AGbar, d\mu)$, take them over to the loop side and
solve them. For the diffeomorphism constraint, most of this task has
been carried out. The operator is well-defined and self-adjoint on the
connection states and can be taken over to the loop side
unambiguously. The result is the expected one: the operator acts on
the loop states $\psi(\a)$ by displacing the loop $\a$ in the argument
via the corresponding diffeomorphism. Its functional form is the one
used in the literature. It is therefore intuitively clear that the
solutions to this constraint functions of (generalized) knot classes.
The question is: What type of functions are we to allow?  As explained
by Petr Haji\v cek (1994), typically, zero is a continuous eigenvalue
of the constraint operators whence elements of the kernel are not
normalizable. Thus, the availability of the Hilbert space by itself
does not provide us with the regularity conditions that must be
satisfied by the {\it physical} states. For this, as Haji\v cek
pointed out, we need a rigged Hilbert space construction. To fix the
ideas, let me recall the situation in the case of a relativistic
particle, where the classical constraint is $P^{\a} P_{\a} +\mu^2 =0$.
In this case, the kinematical Hilbert space can be taken to be
$L^2(\real^4)$.  This space is needed to translate the classical
constraint function to a well-defined operator (whose kernel can then
be found). The operator, of course, is $-\hbar^2 \eta^{\a\b}\d_\a\d_\b
+\mu^2$.  It is self-adjoint on the kinematic Hilbert space
$L^2(\real^4)$.  However, no (non-zero) element in its kernel is
normalizable. These elements belong to the rigged Hilbert space; in
the momentum space, they are distributions with support on the mass
shell. To find out the regularity conditions on the (generalized) knot
invariants, we need to carry out a similar construction and isolate
the appropriate rigged Hilbert space. Work is in progress on this
problem.

For the Hamiltonian constraint, there is still no progress to report
on the mathematical side.  This is not surprising in the light of the
fact that work on the operators $\hat{T}^{ab}(s,t)$ is still in
progress. But the general line of attack is clear. At the heuristic
level, there have been several distinct approaches to the problem of
constructing this operator in the loop representation but they have
all led to equivalent results (Br\"gmann \& Pullin (1993)). In an
optimistic view, this could be taken as a signal that the operator {\it
would} exist rigorously.

To summarize then, there has been a considerable amount of rigorous
work in the last two years and the goal of giving a precise meaning to
the loop transform and the constraint operators as well as that of
extracting the regularity conditions on the physical states now seems
attainable in the near future. The main open problems are: i) extending
the theory to incorporate complex connections; ii) writing the
Hamiltonian constraint as a well-defined operator on $L^2 (\AGbar,
d\mu)$; and, iii) constructing the appropriate rigged Hilbert spaces to
obtain regularity conditions on the physical states.  Work is in
progress on all these problems.

Thus, it appears that the integration theory based on the measure
$d\mu$ will provide the mathematical basis for this non-perturbative
approach to quantum gravity.  However, these techniques may be used
also in other theories of connections which are diffeomorphism
invariant and perhaps even in Yang-Mills theory which is not
diffeomorphism invariant. We saw that the full domain space of quantum
theory, $\AGbar$, can be thought of as the space of homomorphisms from
the full hoop group $\HG$ to the gauge group $SU(2)$. Given a finitely
generated sub-group $S_n$ of the hoop group, we can consider the space
of homomorphisms from it to $SU(2)$. This provides the space $\SU$
which is precisely the domain space of quantum states of a lattice
gauge theory where the lattice is not rectangular but tailored to the
given subgroup $S_n$ of the gauge group.  Thus, what we have is a set
of ``floating lattices,'' each associated with a finitely generated
subgroup of the hoop group. The space $\AGbar$ can be rigorously
recovered as a projective limit of the configuration spaces of lattice
theories (Marolf \& Mour\~ao 1993).  This construction is potentially
quite powerful; it may enable one to take continuum limits of
operators of lattice theories in a completely new fashion. The limit
is obtained not by taking the lattice separation to zero but by
enlarging lattices to probe the continuum connections better and
better, i.e., by considering larger and larger subgroups of the hoop
group. This is a good illustration of the synergestic exchange of ideas
between general relativity and gauge theories that is made possible by
this approach to quantum gravity.

\bigskip\bigskip
\goodbreak
{\bf 4. Emergence of low energy physics}

One of the important features that any non-pertrubative approach to
quantum gravity must have is that, in a suitable limit, it should
reproduce the laboratory scale physics correctly. Does the theory
admit states which can be interpreted as smooth geometries on large
scales? Is there at least an approximate notion of time which is
compatible with a space-time picture? Can one regard non-gravitational
physical fields as evolving unitarily with respect to this time? Is
there an approximate Hamiltonian governing this evolution? One often
takes for granted that the answers to such questions would be
obviously ``yes.''  However, in genuinely non-perturbative treatments,
this is by no means clear a priori; one may be working in a sector of
the quantum theory which does not admit the correct or unambiguous
classical limit.  For example, the sector may correspond to a confined
phase which has no classical analog or the limit may yield a wrong
number even for the macroscopic dimensions of space-time!

In this section, I will present two examples of such results that have
been obtained within the present non-perturbative framework which were
not discussed by other speakers.  The first involves the existence of
quantum states which approximate classical geometries at macroscopic
scales (Ashtekar et al, 1992).  I will begin by showing that certain
operators representing {\it geometrical observables} can be regulated
in a way that respects the diffeomorphism invariance of the underlying
theory.  What is more, these regulated operators are finite {\it
without any renormalization}. Using these operators, one can ask if
there exist loop states which approximate smooth geometry at large
scales. Not only is the answer in the affirmative but, furthermore,
{\it these states exhibit a discrete structure of a definite type at
the Planck scale}. (For further details, see, e.g., Rovelli \& Smolin
(1990), Ashtekar (1992) and Smolin (1993).)  The second example
involves the issue of time in a certain weak field truncation of the
theory.  Here, I will only summarize overall the situation,
emphasizing those points which will be needed in the next section (For
details, see, Ashtekar (1991), chapter 12 and, for more recent
developments, Rovelli and Smolin (1993).)

{\sl 4.1 Non-perturbative regularization}

Let us begin with the issue of regularization. As noted by Guilliani
(1994), in the present framework, the spatial metric is constructed
from products of ``electric fields'' $E^a_i$. It is thus a
``composite'' field given by ${q}^{ab}(x) ={E}^{ai}(x){E}^b_i(x)$.
(This field is, as emphasized by other speakers, a density of weight
2. To keep the notation simple, however, I will drop the tildes that
are generally used to denote the density weights.)  In the quantum
theory, therefore, this operator must be regulated. The obvious
possibility is point splitting. One might set $q^{ab}(x) = \lim_{y\to
x} {E}^{ai}(x) {E}^b_i(y)$.  However, the procedure violates gauge
invariance since the internal indices at two {\it different} points
have been contracted. There is, however, a suitable modification that
will ensure gauge invariance. Consider the field $T^{aa'}[\a] (y',y)$,
labelled by a closed loop $\a$ and points $y$ and $y'$ thereon,
defined in the classical theory by:
$$T^{aa'}[\a ] (y,y') := {1\over 2} \Tr\big[({\cal P} \exp\-
G\int_{y'}^y A_b dl^b) {E}^a(y')\- ({\cal P}\exp \- G\int_y^{y'}A_c
dl^c)\- {E}^{a'}(y)\big]. \eqno(4.1)$$
In the limit $\a$ shrinks to zero, $T^{aa'}[\a](y,y')$ tends to $-4
q^{aa'}$.

Now, in quantum theory, one can define the action of the operator
$\hat{T}^{aa'}[\a](y,y')$ directly on the loop states $\psi(\b)$.  The
explicit form will not be needed here. We only note that using the
bra-ket notation, $\psi(\b) = \IP{\b}{\psi}$ the action can be
specified easily. Indeed, $\bra{\b}\circ\hat{T}^{aa'}[\a](y,y')$ is
rather simple: if a loop $\beta$ does not intersect $\a$ at $y$ or
$y'$, the operator simply annihilates the bra $\langle\beta\mid$ while
if an intersection does occur, it breaks and re-routes the loop
$\beta$, each routing being assigned a specific weight. One may
therefore try to define a quantum operator $\hat{q}^{aa'}$ as a limit
of $\hat{T}^{aa'}[\a ]$ as $\a$ shrinks to zero. The resulting
operator does exist after suitable regularization {\it and}
renormalization. However (because of the density weights involved) the
operator necessarily carries a memory of the background metric used in
regularization. Thus, the idea of defining the metric operator again
fails.  In fact one can give general qualitative arguments to say that
there are no {\it local}, operators which carry the metric information
{\it and} which are independent of background fields (used in the
regularization). Thus, in quantum theory, the absence of background
fields introduces new difficulties.  That such difficulties would
arise was recognized quite early by Chris Isham and John Klauder.

There do exist, however, {\it non-local} operators which can be
regulated in a way that respects diffeomorphism invariance.

As the first example, consider the function $Q(\omega)$ --representing
the smeared 3-metric-- on the classical phase space, defined by
$$Q(\omega):= \int d^3x\ \ (q^{ab}\omega_a\omega_b)^{1\over 2}\- ,
\eqno(4.2)$$
where $\omega_a$ is any smooth 1-form of compact support.  (Note that
the integral is well-defined without the need of a background volume
element because $q^{ab}$ is a density of weight 2.)  It is important
to emphasize that, in spite of the notation, $Q(\omega)$ is {\it not}
obtained by smearing a {\it distribution} with a test field; because
of the square-root, $Q(\omega)$ is {\it not} linear in $\omega$. We
can, nonetheless define the corresponding quantum operator as follows.
Let us choose on $\Sigma$ test fields $f_\epsilon (x,y)$ (which are
densities of weight one in $x$ and) which satisfy:
$$\lim_{\epsilon\to 0} \int_\Sigma \- d^3x\-\- f_\epsilon(x,y)\- g(x)
= g(y) \eqno(4.3)$$
for all smooth functions of compact support $g(x)$. If $\Sigma$ is
topologically $\real^3$, for example, we can construct these test
fields as follows:
$$f_\epsilon (x,y) = {\sqrt{h(x)}\over {\pi^{3\over 2}\epsilon^3}}
\-\- \exp -{\mid \vec x -\vec y\mid^2\over 2\epsilon^2},
\eqno(4.4)$$
where $\vec x$ are the Cartesian coordinates labeling the point $x$
and $h(x)$ is a ``background'' scalar density of weight 2. Next, let
us define
$$ q^{aa'}_\epsilon(x) = -{1\over 4}\int_\Sigma d^3y \int_\Sigma
d^3y' f_\epsilon(x,y) f_\epsilon(x,y') T^{aa'}(y,y').
\eqno(4.5)$$
As $\epsilon$ tends to zero, the right side tends to $q^{ab}$ because
the test fields force both the points $y$ and $y'$ to approach $x$,
and hence the loop passing through $y, y'$, used in the definition of
$T^{aa'}(y,y')$, to zero. It is now tempting to try to define a local
metric operator $\hat{q}^{aa'}$ corresponding to $q^{aa'}$ by
replacing $T^{aa'}(y,y')$ in (4.5) by its quantum analog and then
taking the limit. One finds that the limit does exist provided we
first renormalize $\hat{q}_\epsilon^{aa'}$ by an appropriate power of
$\epsilon$. However, as before, the answer depends on the background
structure (such as the density $h(x)$) used to construct the test
fields $f_\epsilon (x,y)$. If, however, one tries to construct the
quantum analog of the {\it non-local} classical variable $Q(\omega)$,
this problem disappears. To see this, let us first express $Q(\omega)$
using (4.5) as:
$$ Q(\omega) = \lim_{\epsilon\to 0} \int_\Sigma \- d^3x \-\-
(q_\epsilon^{aa'}\omega_a\omega_{a'})^{1\over 2}.
\eqno(4.6)$$
The required quantum operator $\hat{Q}(\omega)$ on the loop states
can now be obtained by replacing $T^{aa'}(y,y')$ by the operator
$\hat{T}^{aa'}(y,y')$. A careful calculation shows that: i) the
resulting operator {\it is} well-defined on loop states; ii) no
renormalization is necessary, i.e., the limit is automatically {\it
finite}; and, iii) the final answer carries no imprint of the
background structure (such as the density $h(x)$ or, more generally,
the specific choice of the test fields $f_\epsilon (x,y)$) used in
regularization. To write out its explicit expression, let me restrict
myself to smooth loops $\a$ without any self-intersection. Then, the
action is given simply by:
$$ \langle\a\mid\circ\hat{Q}(\omega ) = \l_P^2\ \ \oint_\a ds
|\dot{\a}^a \omega_a|\>\cdot \langle \a\mid , \eqno(4.7)$$
where $l_P= \sqrt{G\hbar}$ is the Planck length, $s$, a parameter
along the loop and $\dot{\a}^a$ the tangent vector to the loop. In
this calculation, the operation of taking the square-root is
straightforward because the relevant operators are diagonal in the
loop representation.  This is analogous to the fact that, in the
position representation of non-relativistic quantum mechanics, we can
set $<x|\circ\sqrt{\hat{X}^2} = <x|\cdot |x|$
without recourse to the detailed spectral theory. The $G$ in $l_P$ of
(4.7) comes from the fact that $GA_a^i$ has the usual dimensions of a
connection while $\hbar$ comes from the fact that $\hat{E}^a_i$ is
$\hbar$ times a functional derivative. The final result is that, on
non-intersecting loops, the operator acts simply by multiplication:
the loop representation is well-suited to find states in which the
3-geometry --rather than its time evolution-- is sharp.

The second class of operators corresponds to the area of 2-surfaces.
Note first that, given a smooth 2-surface S in $\Sigma$, its area
${\cal A}_S$ is a function on the classical phase space. We first
express it using the classical loop variables. Let us divide the
surface $S$ into a large number $N$ of area elements $S_I, I=1,2...N$,
and set ${\cal A}_I^{\rm appr}$ to be
$$ {\cal A}_I^{\rm appr} = -{1\over 4}\left[ \int_{S_I} d^2S^{bc}(x)\-\-
\eta_{abc}\int_{{\cal S}_I} d^2S^{ b'c'} (x')\-\- \eta_{a'b'c'}\,
   T^{aa'}(x,x') \right]^{1\over 2}, \eqno(4.8)$$
where $\eta_{abc}$ is the (metric independent) Levi-Civita density of
weight $-1$. It is easy to show that ${\cal A}_I^{\rm appr}$
approximates the area function (on the phase space) defined by the
surface elements $S_I$, the approximation becoming better as $S_I$
--and hence loops with points  $x$ and $x'$ used in the definition of
$T^{aa'}$-- shrink.  Therefore, the total area ${\cal A}_S$ associated
with $S$ is given by
$$ {\cal A}_S = \lim_{N \rightarrow\infty}\, \, \sum_{I=1}^{N} \-
   {\cal A}_I^{\rm appr}. \eqno(4.9)$$
To obtain the quantum operator $\hat{\cal A}_S$, we simply replace
$T^{aa'}$ in (4.8) by the quantum loop operator $\hat{T}^{aa'}$.
This somewhat indirect procedure is necessary because, as indicated
above, there is no well-defined operator-valued distribution that
represents the metric or its area element {\it at a point}. Again, the
operator $\hat{\cal A}_S$ turns out to be finite. Its action,
evaluated on a nonintersecting loop $\a$ (for simplicity), is given
by:
$$  \langle \a | \circ \hat{\cal A}_S  =
   {l_p^2\over 2} \, \> I(S,\a )\- \cdot \langle\a | , \eqno(4.10)$$
where $I(S,\a)$ is simply the {\it unoriented} intersection number
between the 2-surface $S$ and the loop $\alpha$. (One obtains the {\it
un}oriented intersection number here and the absolute sign in the
integrand of (4.7) because of the square-root operation involved in
the definition of these operators.) Thus, in essence, ``a loop $\a$
contributes half a Planck unit of area to any surface it intersects.''

The fact that the area operator also acts simply by multiplication on
non-intersecting loops lends further support to the idea that the loop
representation is well-suited to ``diagonalize'' operators describing
the 3-geometry. Indeed, we can immediately construct a large set of
simultaneous eigenbras of the smeared metric and the area operators.
There is one, $\langle\a|$, associated to every nonintersecting loop
$\a$. Note that the corresponding eigenvalues of area are {\it
quantized} in integral multiples of $l_P^2/2$.  There are also
eigenstates associated with intersecting loops which, however, I will
not go into to since the discussion quickly becomes rather involved
technically.

\bigskip\goodbreak
{\sl 4.2 Weaves}

With these operators on hand, we can now turn to the construction of
quantum loop states that approximate the classical metric $h_{ab}$ on
$\S$ on a scale large compared to the Planck length.  The basic idea
is to weave the classical metric out of quantum loops by spacing them
so that, on an average, precisely one line crosses {\it any} surface
element whose area, {\it as measured by the given} $h_{ab}$ is one
Planck unit. Such loop states will be called {\it weaves}. Note that
these states are not uniquely picked out since our requirement is
rather weak.  Indeed, given a weave approximating a specific classical
metric, one can obtain others, approximating the same classical
metric.

Let us begin with a concrete example of such a state which will
approximate a {\it flat} metric $h_{ab}$.  To construct this state, we
proceed as follows. Using this metric, let us introduce a random
distribution of points on $\Sigma = \real^3$ with density $n$ (so that
in any given volume $V$ there are $nV(1+ {\cal O}(1/\sqrt{nV}))$
points). Center a circle of radius $a = (1/n)^{1\over 3}$ at each of
these points, with a random orientation.  We assume that $a<< L$, so
that there is a large number of (non-intersecting but, generically,
{\it linked}) loops in a macroscopic volume $L^3$. Denote the
collection of these circles by $\Delta_a$. As noted in section 4, due
to trace identities, products of Wilson loop functionals $T_{\at}$ can
be expressed as linear combinations of Wilson loop functionals. As a
consequence, it turns out that the bras defined by multi-loops are
equivalent to linear combinations of single loop bras. Therefore, for
each choice of the parameter $a$, there is a well-defined bra
$\langle\Delta_a|$. This is our candidate weave state.

Let us consider the observable $\hat{Q}[\omega]$. To see if
$\langle\Delta_a|$ reproduces the geometry determined by the classical
metric $h_{ab}$ on a scale $L>>l_p$, let us introduce a 1-form
$\omega_a$ which is {\it slowly varying on the scale} $L$ and compare
the value $Q[\omega](h)$ of the classical $Q[\omega]$ evaluated at the
metric $h_{ab}$, with the action of the quantum operator
$\hat{Q}[\omega]$ on $\langle\Delta_a|$. A detailed calculation yields:
$$ \langle\Delta_a|\circ \hat{Q}[\omega] =  \left[{\pi\over 2} \- \
({l_p\over a})^2 \, Q[w](h) + {\cal O}({a\over L})\right]\-  \cdot
\langle\Delta_a|. \eqno(4.11)$$
Thus, $\langle\Delta_a|$ is an eigenstate of $\hat{Q}[\omega ]$ and
the corresponding eigenvalue is closely related to $Q[\omega](h)$.
However, even to the leading order, the two are unequal {\it unless}
the parameter $a$ --the average distance between the centers of
loops-- {\it equals} $\sqrt{\pi/2}\, l_p$. More precisely, (4.11)
can be interpreted as follows. Let us write the leading coefficient on
the right side of this equation as $(1/4)(2\pi a/l_p)(nl_p^3)$.  Since
this has to be unity for the weave to reproduce the classical value
(to leading order), we see that $\Delta_a$ should contain, on an
average, one fourth Planck length of curve per Planck volume, where
lengths and volumes are measured using $h_{ab}$.

The situation is the same for the area operators $\hat{\cal A}_S$. Let
$S$ be a 2-surface whose extrinsic curvature varies slowly on a scale
$L >>l_P$. One can evaluate the action of the area operator on
$\langle \Delta_a|$ and compare the eigenvalue obtained with the value
of the area assigned to $S$ by the given flat metric $h_{ab}$. Again,
the eigenvalue can be re-expressed as a sum of two terms, the leading
term which has the desired form, except for an overall coefficient
which depends on the mean separation $a$ of loops constituting
$\Delta_a$, and a correction term which is of the order of ${\cal
O}({a\over L})$. We require that the coefficient be so adjusted that
the leading term agrees with the classical result. This occurs, again,
precisely when $a = \sqrt{\pi/2}\- l_p$. It is interesting to note
that the details of the calculations which enable one to express the
eigenvalues in terms of the mean separation are rather different for
the two observables. In spite of this, the final constraint on the
mean separation is {\it precisely} the same.

Let us explore the meaning and implications of these results.
\item{1)} {It is generally accepted that, to obtain
classical behavior from quantum theory, one needs two things: i) an
appropriate coarse graining, and, ii) special states. In our
procedure, the slowly varying test fields $\omega_a$ and surfaces $S$
with slowly varying extrinsic curvature enable us to perform the
appropriate coarse graining while weaves --with the precisely tuned
mean separation $a$-- are the special states. There is, however,
something rather startling: The restriction on the mean separation $a$
--i.e., on the {\it short distance} behavior of the multi-loop
$\Delta_a$-- came from the requirement that $\langle\Delta_a|$ should
approximate the classical metric $h_{ab}$ on {\it large scales} $L$!}
\item{2)} {In the limit $a\to\infty$, the eigenvalues of the two
operators on $\langle\Delta_a|$ go to zero. This is not too surprising.
Roughly, in a state represented by any loop $\a$, one expects the
quantum geometry to be excited just at the points of the loops. If the
loops are {\it very} far away from each other as measured by the
fiducial $h_{ab}$, there would be macroscopic regions devoid of
excitations where the quantum geometry would seem to correspond to a
zero metric.}
\item{3)} {The result of the opposite limit,  however, {\it is}
surprising.  One might have naively expected that the best
approximation to the classical metric would occur in the continuum
limit in which the separation $a$ between loops goes to zero. However,
the explicit calculation outlined above shows that this is not the
case: as $a$ tends to zero, the leading terms in the eigenvalues of
$\hat{Q}[\omega]$ and ${\cal A}_S$ actually diverge!  (One's first
impulse from lattice gauge theories may be to say that the limit is
divergent simply because we are not rescaling, i.e., renormalizing the
operator appropriately. Note, however, that, in contrast to the
calculations one performs in lattice theories, here, we {\it already}
have a well defined operator in the continuum. We are only probing the
properties of its eigenvectors and eigenvalues, whence there is
nothing to renormalize.)  It is, however, easy to see the reason
underlying this behavior. Intuitively, the factors of the Planck
length in (4.7) and (4.10) force each loop in the weave to
contribute a Planck unit to the eigenvalue of the two geometrical
observables.  In the limit $a\to 0$, the number of loops in any fixed
volume (relative to the fiducial $h_{ab}$) grows unboundedly and the
eigenvalue diverges.}
\item{4)} {It is important to note the structure of the argument. In
non-perturbative quantum gravity, there is no background space-time.
Hence, terms such as ``slowly varying'' or ``microscopic'' or
``macroscopic'' have, a priori, no physical meaning. One must do some
extra work, introduce some extra structure to make them meaningful.
The required structure should come from the very questions one wants
to ask. Here, the questions had to do with approximating a classical
geometry. Therefore, we could {\it begin} with classical metric
$h_{ab}$. We used it repeatedly in the construction: to introduce the
length scale $L$, to speak of ``slowly varying'' fields $\omega_a$ and
surfaces $S$, and, to construct the weave itself. The final result is
then a consistency argument: If we construct the weave according to
the given prescription, then we find that it approximates $h_{ab}$ on
macroscopic scales $L$ provided we choose the mean separation $a$ to
be $\sqrt{\pi/2} l_p$, where all lengths are measured relative to the
same $h_{ab}$.}
\item{5)} {Note that there is a considerable non-uniqueness in the
construction. As we noted already, a given 3-geometry can lead to
distinct weave states; our construction only serves to make the
existence of such states explicit. For example, there is no reason to
fix the radius $r$ of the individual loops to be $a$. For the
calculation to work, we only need to ensure that the loops are large
enough so that they are generically linked and small enough so that
the values of the slowly varying fields on each loop can be regarded
as constants plus error terms which we can afford to keep in the final
expression. Thus, it is easy to obtain a 2-parameter family of weave
states, parametrized by $r$ and $a$. The condition that the leading
order terms reproduce the classical values determined by $h_{ab}$ then
gives a relation between $r$, $a$ and $l_P$ which again implies
discreteness.  Clearly, one can further enlarge this freedom
considerably: There are a lot of eigenbras of the smeared-metric
and the area operators whose eigenvalues approximate the classical
values determined by $h_{ab}$ up to terms of the order ${\cal O}({l_p
\over L})$ since this approximation ignores Planck scale quantum
fluctuations. }
\item{6)} {Finally, I would like to emphasize that, at a conceptual
level, the important point is that the eigenvalues of $\hat{Q}[\omega
]$ and ${\cal A}[S]$ can be {\it discrete}.}

Let me conclude the discussion on weaves with two remarks. First, it
is not difficult to extend the above construction to obtain weave
states for curved metrics $g_{ab}$ which are slowly varying with
respect to a flat metric $h_{ab}$. Given such a metric, one can find a
slowly varying tensor field $t_a{}^b$, such that the metric $g_{ab}$
can be expressed as $t_a{}^c t_b{}^d h_{cd}$. Then, given a weave of
the type $\langle\Delta|$ considered above approximating $h_{ab}$, we
can ``deform'' each circle in the multi-loop $\Delta$ using $t_a{}^b$
to obtain a new weave $\langle \Delta |_{t}$ which approximates
$g_{ab}$ in the same sense as $\langle\Delta|$ approximates $h_{ab}$.
(See, also, Zegwaard (1993, 1994) for the weave corresponding to the
Schwarzschild black-hole.)  The second remark is that since the
weaves are eigenbras of the operators that capture the 3-geometry,
they do not approximate 4-geometries. To obtain a state that can
approximate Minkowski space-time, for example, one has to consider a
loop state that resembles a ``coherent state'' peaked at the weave
$\Delta_a$. In that state, neither the 3-geometry nor the
time-derivative thereof would be sharp; but they would have minimum
spreads allowed by the uncertainty principle. This issue has been
examined in detail by Iwasaki and Rovelli (1993).

Since these results are both unexpected and interesting, it is
important to probe their origin. We see no analogous results in
familiar theories. For example, the eigenvalues of the fluxes of
electric or magnetic fields are not quantized in QED nor do the
linearized analogs of our geometric operators admit discrete
eigenvalues in spin-2 gravity. Why then did we find qualitatively
different results? The technical answer is simply that the familiar
results refer to the {\it Fock representation} for photons and
gravitons while we are using a completely different representation
here. Thus, the results {\it are} tied to our specific choice of the
representation. Why do we not use Fock or Fock-like states? It is not
because we insist on working with loops rather than space-time fields
such as connections. Indeed, one {\it can} translate the Fock
representation of gravitons and photons to the loop picture. (See,
e.g., Ashtekar et al (1991) and Ashtekar \& Rovelli, (1992).)  And
then, as in the Fock space, the discrete structures of the type we
found in this section simply disappear. However, to construct these
loop representations, one must use a flat background metric and
essentially every step in the construction violates diffeomorphism
invariance. Indeed, there is simply no way to construct ``familiar,
Fock-like'' representations without spoiling the diffeomorphism
invariance. Thus, the results we found are, in a sense, a direct
consequence of our desire to carry out a genuinely non-perturbative
quantization without introducing any background structure.  However,
we do not have a uniqueness theorem singling out the measure $d\mu$
which was used to define the loop transform and hence to construct the
loop representation used here. There do exist other diffeomorphism
invariant measures which will lead to other loop representations. The
measure we have used is the most natural and the simplest among the
known ones.  Whether the results presented here depend sensitively on
the choice of the measure is not known.  Therefore, it would be highly
desirable to have a uniqueness theorem which tells us that, on
physical grounds, we should restrict ourselves to a specific (class
of) measure(s).

My overall viewpoint is that one should simultaneously proceed along
two lines: i) one should take these results as an indication that we
are on the right track and push heuristic calculations within this
general framework; and, ii) one should try to put the available
heuristic results on rigorous mathematical footing to avoid the danger
of ``wandering off'' in unsound directions.

\bigskip\goodbreak
{\sl 4.3 The issue of time}

The results reported in the last two sub-sections were kinematical;
constraints were not yet imposed. This is because our questions
themselves referred to 3-metrics which fail to be Dirac observables;
in the classical analysis, 3-metrics are well-defined only on the full
(and hence kinematical) phase space rather than on the reduced phase
space (the space of physical or ``true'' degrees of freedom). On the
other hand, as Haji\v cek (1994) explained in his lectures, the
issue of time is {\it not} kinematical; it is intimately intertwined
with the scalar or the Hamiltonian constraint.  Therefore, we must now
bring in this constraint. We will show that in a certain truncation
of the full quantum theory, the issue of time can be resolved
satisfactorily in the asymptotically flat context.

It turns out that, for technical reasons sketched below, (in pure
gravity) it is the connection representation that seems best-suited to
handle this issue.  Thus, the question we wish to now ask is the
following: can we single out a component of the connection, ${}^T\!A$
which can serve as ``internal'' time?  More precisely, can we
re-interpret the scalar --or Hamiltonian-- constraint as telling us
how the ``true'' or the dynamical degrees of freedom ${}^D\!A$ evolve
with respect to the internal clock ${}^T\!A$? The questions themselves
are not new. They have been asked in the context of geometrodynamics
since the sixties.  It turned out, however, that one can not isolate
time in this manner in the metric representation of geometrodynamics
even in the weak field truncation (Kucha\v r (1970)). Why does one then
hope that the situation would be any better in connection dynamics?

Let us begin by analyzing the source of the difficulty. Consider,
first, a parametrized, non-relativistic system. Let the phase space
coordinates be $(q^1, ...q^n; p_1, ...p_n)$ and let the constraint be:
   $$p_1 + H(q^1, q^2,..., q^n; p_2, ... p_n) = 0.\eqno(4.12)$$
Now, if in the quantum theory, we used the $q$-representation, the
quantum constraint has the same form as the time-dependent
Schr\"odinger equation, $i\hbar\d\Psi(q^i)/\d q^1=
\hat{H}\circ\Psi(q^i)$, where $q^1$ plays the role of time. On the
other hand, the momentum representation is equally viable
mathematically. The quantum constraint would now take the form
$\hat{p}^1\Psi(p^i) = -\hat{H}\circ \Psi(p^i)$; it does {\it not}
resemble the time-dependent Schr\"odinger equation. In particular, it
is now hard to isolate time from among the configuration variables on
which the wave function depends. Thus, although two representation may
be mathematically equivalent, one may be better suited to deal with
the issue of time. Since the Dirac transformation theory enable us to
pass from one to another, the issue here is that of convenience.
However, since so many of the issues related to the interpretation of
the framework hinge on the availability of a (possibly approximate)
time variable, a representation which is well-suited to extract an
internal clock has central importance. Let us now return to gravity.
Recall from Beig's lectures that, in the asymptotically flat case, the
Hamiltonian of geometrodynamics can be written as $H_N(q,p) = C_N(q,p)
+ {\cal E}(q)$, where $N$ is any lapse which tends to 1 at infinity at
an appropriate rate and ${\cal E}(q)$ is the ADM energy integral which
depends only on the 3-metric $q_{ab}$ but not on the conjugate
momentum $p^{ab}$. Hence, the scalar constraint corresponding to this
lapse can be written in the form:
 $${\cal E}(q) - H_N(q,p)= 0,\eqno(4.13)$$
which is identical in form to (4.12), with $-{\cal E}(q)$ playing the
role of $p_1$.  The similarity suggests that it is the variable that
is ``conjugate'' to $-{\cal E}(q)$ that should play the role of time.
This idea is attractive also because ${\cal E}(q)$ has the
interpretation of the total energy of the system.  Furthermore, the
structure of $(4.13)$ is unaffected by the presence of matter sources;
it is universal. However, the similarity also suggests that the metric
representation is not likely to be well-suited to extract time from
among the gravitational degrees of freedom. The connection
representation would be better suited since ${\cal E}(q)$ would be a
differential operator in the $A$-representation.  We will see below
that the expectation is borne out in the weak field truncation of the
theory. Note finally that this argument also explains why it is
unlikely that the gravitational loop variables would be well-suited to
extract time: as we saw in the last two sub-sections, the loop
representation is diagonal in the operators carrying information of
the 3-geometry%
\footnote{${}^\dagger$}{If we couple gravity to matter, however, the
picture can change. It is then possible that one of the matter
variables can play the role of an internal clock. This is an
attractive strategy especially in the spatially compact case.}.
With these general remarks out of the way, we can now explain the
precise sense in which the scalar constraint equation can be
interpreted as a time dependent Schr\"odinger equation.

We will now assume that the underlying 2-manifold $\S$ is
topologically $\real^3$. Let us begin by introducing a background
point in the phase space, $(\Az=0, \Ez)$, where $\Ez$ is a flat triad
and expand out the fields $\hat{A}_a^i$ and $\hat{E}^a_i$ that appear
in the constraint operators, in the powers of the deviations
$\hat{A}_a^i-0$ and $\hat{E}^a_i - \Ez$. Since we now have access to a
background triad, $\Ez$, it will be convenient to convert the internal
indices to vector indices on all dynamical fields. Thus, quantum
states can now be regarded as functionals of $A_{ab} := A_a^i
{}^o\!E_{ib}$. Next, using the flat metric ${}^o\!q_{ab}$ obtained
from $\Ez$, we can decompose the symmetric part of $A_{ab}$ into its
transverse-traceless, longitudinal and trace parts. Let us now impose
the quantum constraints order by order in the deviation. The first
order equations imply that the wave functions must depend only on the
symmetric, transverse-traceless parts. That is, all other parts of
$\hat{A}_{ab}$ are at least of second order. Next, imposing the scalar
constraint to the second order yields:
$$ -\textstyle{2\over G}\ (\triangle\ {}^T\!\hat{E})\ \circ \
\Psi({}^D\!A, {}^T\!A) = G \ ({}^{TT}\!\hat{A}_{ab})^\star\ \
({}^{TT}\!\hat{A}^{ab})\ \cdot \ \Psi({}^D\!A, {}^T\!A),
\eqno(4.14)$$
where, $\triangle$ is the Laplacian with respect to ${}^o\!q_{ab}$,
${}^T\!\hat{E}$ is the trace part of $\hat{E}^{ab}$,
${}^{TT}\!\hat{A}_{ab}$ is the (symmetric) transverse traceless part
of $\hat{A}_{ab}$, and where, as before, ${}^T\!A$ and ${}^D\!A$ are
the trace and the dynamical (i.e. all but trace) parts of $A_{ab}$. It
is now natural to introduce the variable $\tau(x)$ which is conjugate
to the operator on the left side of this equation: $\tau(x)=
(-G/2)\triangle^{-1}\cdot {}^T\!A (x)$, to write the equation more
explicitly as:
$$ \hbar\ {\delta \Psi({}^D\!A, {}^T\!A)\over \delta\tau(x)} =
G\ ({}^{TT}\!\hat{A}_{ab})^\star\ \
({}^{TT}\!\hat{A}^{ab})\ \cdot \ \Psi({}^D\!A, {}^T\!A).
\eqno(4.15)$$
Eq (4.15) tells us how to determine the value of the physical state
$\Psi({}^D\!A, {}^T\!A)$ {\it everywhere} on the configuration space
from its value on one ${}^T\!A =\ const.$ hypersurface. In this
sense, the equation can be thought of as a quantum constraint. It
tells us that the physical states are not freely specifiable; their
functional form is constrained. To obtain the ``evolution''
interpretation, it suffices to use just ``one component'' of this
functional differential equation. The idea is simply to integrate the
equation on $\S$ which corresponds to taking its moment with a lapse
$N=1$. We can simplify the left side of the resulting equation using
the fact that the states are all {\it holomorphic} functionals of
$A_{ab}$ and, the right side, by going to the momentum space. The
result is:
$$i\hbar \big(\lint d^3x {\delta\over {\delta{\rm Im}\tau(x)}}\big)\circ
\Psi({}^D\!A, {}^T\!A) = \hbar\big(\int d^3 k |k|\ \ {}^{TT}\!A_{ab}(k)\
{\delta\over{\delta{}^{TT}\!A_{ab}(k)}}\big) \circ \Psi({}^D\!A,{}^T\!A).
\eqno(4.16)$$
Thus, the imaginary part, ${\rm Im}\tau(x)$, of $\tau(x)$ plays the
role of time. The operator on the right side is precisely the
Hamiltonian of the truncated theory. Thus, in the second order
truncation, the scalar constraint reproduces the time-dependent
Schr\"odinger equation. This feature remains intact if one includes
matter fields as sources.

As I have emphasized, in non-perturbative quantum gravity, we do not
have access to a classical space-time. Even in the above truncation
procedure, we worked only with the 3-manifold $\S$ and expanded
operators around some background fields on $\S$. Thus, we did not have
access to a classically defined time variable. Rather, we were able to
isolate, from among the mathematical variables contained in $A_a^i$, a
preferred time $t$ which serves as an ``internal clock'' with respect
to which the wave function evolves. Put differently, by identifying
time in the components of $A_a^i$, we have derived the Schr\"odinger
equation of weak-field gravity without having a direct access to a
space-time metric or even a 4-manifold. After having obtained the
result, we can look back and see that we could have obtained the same
result in a space-time picture. That is, the true dynamics takes place
in the infinite dimensional configuration space of connections.
However, {\it in suitable approximations} that dynamics can be
re-interpreted as taking place in a suitably constructed 4-dimensional
space-time.  There may, however, be instances when the approximation
would break down and no space-time picture is adequate. However, if
any non-perturbative description is to be viable at all, it better be
the case that it reduces to Minkowskian physics with Schr\"odinger
evolution for quantum fields in an appropriate limit. As Kucha\v r
(1970) pointed out almost 25 years ago, this is hard to achieve directly
in geometrodynamics. Connection dynamics, as we have seen, is better
suited for this task.

Finally, there is another approach to the extraction of time: coupling
to matter fields. Recently, Rovelli and Smolin (1993) have made
significant progress by coupling gravity to a scalar and other matter
fields and using the scalar field to define the internal clock. The
remarkable aspect of the development is that the resulting {\it true}
quantum Hamiltonian is a manageable operator in the loop
representation. The precise domain in which time can be so extracted
--and hence, the nature of the implicit approximation scheme-- is,
however, not clear yet. Nonetheless, the fact that concrete calculations
can be done in the full, non-truncated theory, is already quite exciting.
The problem of time is an old one and much effort has gone into trying to
find a clean, exact solution to that problem. These approximation
methods and concrete calculations may well provide the type of insight
that has been missing.

\bigskip\bigskip
\goodbreak
{\bf 5. Outlook}

As articles in this volume illustrate, in recent years, there has been
notable progress on some of the difficult issues in quantum gravity.
In particular, we have seen that connection dynamics offers a unified
mathematical framework for all four fundamental interactions. Not only
do the loop space methods enable us to go back and forth successfully
between gauge theories and gravity but they have also made contact
with to the fertile area of knot theory. On the mathematical side, the
subject has achieved a surprising degree of maturity over the last two
years and one can look forward to further contributions from the
mathematical physics community --the constructive field theorists, the
$C^\star$-algebra community, topologists and knot theorists-- in the
coming years.

As Br\"ugmann discussed in some detail, the program is of course far
from being complete. But the situation in 2+1-dimensional gravity
suggests that the various difficulties we face may be primarily
technical.  With enough work, in the near future we should be able to
establish that either the diverse goals of the program can in fact be
met or that the program encounters {\it specific} unsurmountable
obstacles. The second type of result would also represent concrete
progress. After all, the program is driven by a ``radical
conservatism'' ---it is based on the well established principles of
general relativity and quantum theory and does not begin by guessing
what the Plank scale structure ought to be. Therefore, if it runs into
specific problems, it is likely that these limitations themselves
will suggest the required modifications.

What directions is the program likely to take in the near future?  It
is perhaps of interest to list a few open problems whose solution will
considerably improve the current state of the field.  This is {\it
not} an exhaustive list. My primary goal is to illustrate, with
concrete examples, promising directions in the hope is that these
examples will stimulate young researchers as well as experts in
various areas to contribute to the field:
\item{1)} {\it Mathematical issues:} As we saw in section 3, the level
of mathematical precision in the Rovelli-Smolin loop transform has
improved considerably. However, very little is known regarding the
inverse transform. Can we characterize the loop functionals in the
image of the transform in a simple way without having to refer
to the connection representation? Can we define a measure on the hoop
group and define the inverse transform? Does the equivalent of the
Plancherel theorem --which makes the Fourier transform so powerful--
hold?  A promising direction is being pursued by Gambini and his
collaborators where the hoops group is replaced by a group of
``smoothened out'' hoops (Di Bortolo et al, 1993). This group has the
structure of a Lie group and offers a new approach to the problem of
regularization of various operators. This may well be the framework
needed to make the inverse transform well-defined.
\item{2)} {\it Euclideanization:} In constructive quantum field
theory, the developments which culminated in the Osterwalder-Schrader
system of Euclidean axioms caused a burst of activity. Suddenly, a
variety of powerful techniques became available. The key result was
the demonstration of the equivalence of the Osterwalder-Schrader
system to the Wightman system: if we have a theory satisfying the
first set, there exists a theory satisfying the second, even though it
may be hard in practice to construct it explicitly. Thus, it is still
the Lorentzian theory that is of physical interest; it is just that
the mathematical problem of constructing such a theory can be reduced
to a more tractable problem in the Euclidean framework. Does an
analogous Euclidean system exist for general relativity? We know that
naive attempts at trying to get at the physical, Lorentzian theory via
the obvious Euclidean constructions can not succeed (Mena-Marugan,
1993).  Are there more subtle constructions? The discovery of such a
procedure would be a key contribution since many of the technically
hard problems simplify enormously in the Euclidean domain.
\item{3)} {\it Midisuperspaces:} The midi-superspace of solutions to
(3+1-dimensional) Einstein's equation with one (space-like) Killing
field is an especially fertile area for future work. Mathematically,
this system is equivalent to 2+1-dimensional general relativity
coupled to a non-linear sigma-model. If the Killing field in question
is hypersurface orthogonal, the matter field reduces to a single
scalar field satisfying the wave equation. (If, in addition, the norm
of the Killing field is constant, the matter field disappears and we
have vacuum 2+1 gravity, which we already know how to quantize.) Thus,
in these midi-superspaces, the problem reduces to that of quantizing
2+1 gravity with matter. It is a genuine field theory with an infinite
number of degrees of freedom. However, it has two key simplifications.
First, the reality conditions now become significantly simpler:
through an ``internal Wick-rotation'' along the Killing direction, one
is led to consider {\it real} $SU(1,1)$ connections in place of the
complex $SU(2)$ ones (Ashtekar \& Varadarajan, 1992a). Second, in the
asymptotically flat case, the total Hamiltonian of the system is (not
only non-negative but also) {\it bounded from above} indicating
strongly that the quantum theory may be finite (Ashtekar \&
Varadarajan, 1992b). This midi-superspace thus has just the right
blend of technical simplifications and physical generality to serve as
a valuable stepping stone to the full theory.
\item{4)} {\it Solutions to the Hamiltonian constraint:} The ``smoothened
loop'' techniques introduced by Gambini and his collaborators have led
to interesting new solutions to the Hamiltonian constraints which are
related to some well-known knot invariants (Br\"ugmann et al,
1992a,b). Furthermore, it is clear that these are just the simplest
applications of the framework. These methods are powerful and have
opened up a new line of attack to the problem of solving the
Hamiltonian constraint. Obtaining explicit solutions and understanding
their physical meaning is important because that would strengthen our
intuition considerably. However, it is also highly desirable to get a
handle on the {\it structure} of the space of solutions. To define the
inner-product on physical states, for example, it is not necessary to
have the solutions explicitly. What we need is an understanding of the
various mathematical structures that naturally exist on this space. So
far, this issue has attracted very little attention. Finally, recently
Barbero-Gonzalez (1993) has recast general relativity in 3 and 4
dimensions as a theory of {\it two} connections in which the triad
field is secondary and does not even enter the basic equations. This
is an exciting development with a lot of potential and may, in
particular, open up new avenues to the issue of solving the quantum
constraints.
\item{5)} {\it Approximation methods} To address
physical issues, we need to develop the approximation methods further.
For example, we saw in section 4.3 that by a truncation procedure
around a Minkowskian initial data, one can recover the familiar
laboratory physics from the scalar constraint. An obvious question
then is what would happen if we truncated around a black-hole initial
data. Can one automatically recover the Hawking effect? There are
indications that one would. The analysis of ``embroidery around
weaves'' due to Iwasaki and Rovelli (1993) is likely to play an
important role in this procedure.  The final result will be a
derivation of the Hawking effect from the loop representation of full
quantum gravity (i.e., from ``above'') rather than from quantum field
theory in curved space-times (i.e., from ``below''). A second problem
involves matter couplings. Rovelli and Smolin (1993) have considered
the coupling to a scalar field while Matschull and Nicoli (1993a, b)
and Morales-T\'ecotl and Rovelli (1994) have discussed fermion
coupling.  As I mentioned in section 4.3, Rovelli and Smolin use the
scalar field to define an approximate time variable and thus recover a
time dependent Schr\"odinger equation with a true Hamiltonian. For
spinor field sources, they are developing an approximation technique
to analyze the action of this Hamiltonian. A great deal of physical
understanding is expected to come from such analyses of matter
couplings.
\smallskip

Different people, of course, have different expectations of quantum
gravity. To conclude, I would like to ask your indulgence for a moment
or two as I express my personal views and prejudices.

The aim of the canonical program, as I see it, is to find out if a
consistent quantum version of general relativity coupled to matter
fields can exist non-perturbatively. Thus, one focusess on certain
issues and, at least at the first go, ignores others. The focus is on
quantum geometry, on diffeomorphism invariance, on internal clocks, on
mathematical methods needed in any non-perturbative treatment of
gravity, and, although these were not discussed at this workshop, on
new approximation methods. The program is certainly far from being
complete. However, it has not run into an unsurmountable obstacle
either. The basic spirit, I believe, is the same as the one that drove
physicists when they were groping in the dark during the development
of quantum mechanics. One pushes the promising ideas and the plausible
techniques as hard as one can. Either they keep working or break down
at some point. If they break down, one focuses on where and why they fail.
Occasionally, the breakdown suggests brand new strategies and changes
the direction of the effort substantially.  We know that, during the
development of quantum mechanics, most ideas did not work and, at any
given time, almost everyone was barking up the wrong tree. And yet
this strategy of pushing the ideas ruthlessly as far as they can go
always produced interesting results. In the same spirit, the hope now
is that we will learn something interesting and perhaps even something
deep.  The evolution from metrics to connections to loops to weaves
and knots represents progress along these lines.

Even if the program succeeds and produces a consistent,
non-perturbative theory, there is no guarantee, of course, that this
theory would be the correct one. Indeed, to make concrete progress,
some issues were put aside at first and, in the final analysis, these
may well turn out to be so central to the problem that they have to be
incorporated right at the start. For example, it may well be, as the
majority of the particle physics community holds, that real progress
would not occur unless one has, from the very beginning, a principle
which unifies the dynamics of all interactions. It may be
inappropriate to seek insights into the quantum nature of geometry
without incorporating {\it all} excitations of the superstring. It may
also be, as several relativists believe, that real progress would not
occur unless the approach is geared to tackle the measurement problems
of quantum mechanics from the start; unless the mathematical structure
underlying quantum mechanics is made to absorb some fundamental
non-linearity at the outset. Most people working on this canonical
approach are sympathetic to these ideas in a general way.  Indeed,
work is in progress on both of these frontiers. However, the general
sentiment is that the approach need not be {\it based} on such
premises. Once the subject has evolved sufficiently, new ideas will
come up to tackle such issues.  There is often a tendency to
underestimate the value of having a consistent, non-perturbative
theory irrespective of whether it ultimately turns out to be the
correct one physically. There appears to be an overriding sentiment
that a quantum theory of gravity {\it must} solve all sorts of
problems including those that are not, at least in any obvious way,
intrinsic to gravity.  And when one lists all these issues and all the
associated problems, one is often so struck by the enormity of the
task that a sense of hopelessness seems to take over. For a long time
now, I think, there has been an undercurrent of pessimism at least in
some parts of the relativity community: One does what one can but
secretly (or, sometimes, openly!) believes that the task is {\it way}
beyond us.  This may well be the case. But I think it is also not
obvious that this is really the case. Perhaps we should not try to
solve all problems at once. We can suitably restrict our goals and
pursue these ``modest'' programs with full enthusiasm and hope of
success. I think we need a more outgoing attitude here, more
aggressive spirits and more optimistic hearts!

\bigskip\bigskip
\goodbreak

{\bf Acknowledgments}

I benefited a great deal from the talks as well as informal
discussions with the participants of this workshop. My special thanks
go to Helmut Friedrich for organizing this stimulating meeting and for
his patience with all administrative matters. This work was supported
in part by the NSF grant PHY93-96246 and by the Eberly research fund
of the Pennsylvania State University.

\bigskip\bigskip
\goodbreak
{\bf References}

\item{}{Achucarro, A. \& Townsend, P.  (1986) Phys. Lett. {\bf B180},
85-89.}
\item{}{Agishtein, A. \& Migdal, A.  (1992) Mod. Phys. Lett.
{\bf 7}, 1039-1061.}
\item{}{Amati, D. Ciafaloni, M. \& Veneziano, G.  (1990) Nucl.
Phys. {\bf B347}, 550-590.}
\item{}{Aspinwall, P. S.  (1993) Mirror symmetry, talk at the 1993
Rutherford-Appleton Meeting.}
\item{}{Ashtekar, A.  (1991) {\it Non-perturbative canonical
gravity}, World Scientific, Singapore.}
\item{}{Ashtekar, A.  (1992) Mathematical problems of
non-perturbative quantum general relativity, SU-GP-92/11-2; to appear
in the proceedings of the 1992 Les Houches summer school, B. Julia
(ed), Elsevier.}
\item{}{Ashtekar, A., Husain, V., Rovelli, C., Samuel, J. \&
Smolin, L.  (1989) Class. Quan. Grav. {\bf 6}, L185-L193.}
\item{}{Ashtekar, A.,  Rovelli, C. \& Smolin, L.  (1991) Phys. Rev.
{\bf D44}, 1740-1755.}
\item{}{Ashtekar, A. \& Isham, C.  (1992) Class. Quan. Grav. {\bf 9},
1069-1100.}
\item{}{Ashtekar, A. \& Rovelli, C.  (1992) Class. Quan. Grav. {\bf 91}
1121-1150.}
\item{}{Ashtekar, A., Rovelli, C. \& Smolin L.  (1992) Phys. Rev.
Lett. {\bf 69} 237-240.}
\item{}{Ashtekar, A. \& Varadarajan, M.  (1992a,b) in preparation.}
\item{}{Ashtekar, A. \& Lewandowski, J.  (1993a) Representation theory of
analytic holonomy $C^\star$ algebras; to appear in {\it Knots and
quantum gravity}, J. Baez (ed), Oxford University press.}
\item{}{Ashtekar A. \& Lewandowski, J.  (1993b) Differential calculus
on $\AG$, in preparation.}
\item{}{Ashtekar, A. \& Loll, R. (1994) New loop representations for
2+1-dimensional gravity.}
\item{}{Baez, J.  (1993a) Diffeomorphism-invariant generalized
measures on the space of connections modulo gauge transformations,
hep-th/9305045; to appear in the Proceedings of conference on quantum
topology, L. Crane and D. Yetter (eds).}
\item{}{Baez J.  (1993b) Link invariants, functional integration and
holonomy algebras, hep-th/9301063.}
\item{}{Barbero-Gonzalez, J. F.  (1993) General relativity as a theory
of two connections, CGPG-93/9-5; Int. J. Mod. Phys. {\bf D}(in
press).}
\item{}{Beig, R.  (1994) article in this volume.}
\item{}{Br\"ugmann, B.  (1994) article in this volume.}
\item{}{Br\"ugmann, B., Gambini, R. \& Pullin, J.  (1992a) Phys. Rev. Lett.
{\bf 68}, 431-434.}
\item{}{Br\"ugmann, B., Gambini, R. \& Pullin, J.  (1992b) In:
{\it XXth International conference on differential geometric methods
in physics}, S. Cotta \& A. Rocha (eds), World Scientific, Singapore.}
\item{}{Br\"ugmann, B. \& Marinari, E.  (1993) Phys. Rev. Lett. {\bf 70}
1908-1911.}
\item{}{Br\"ugmann, B. \& Pullin, J.  (1993) Nucl. Phys. {\bf 390}
399-438.}
\item{}{Carlip, S.  (1990) Phys. Rev. {\bf D42}, 2647-2654.}
\item{}{Carlip, S.  (1993) Six ways to quantize (2+1)-dimensional
gravity, gr-qc/9305020.}
\item{}{de Wit, B., Matschull, H.J. \& Nicolai, H. (1993) Phys. Lett.
{\bf B318}, 115.}
\item{}{Di Bortolo, R. Gambini \& Griego, J. (1992) Comm. Math. Phys.
(to appear)}
\item{}{Gambini, R. \& Trias A. (1986) Nucl. Phys. {\bf B278}, 436-448.}
\item{}{Gross, D. \& Mende, P (1988) Nucl. Phys. {\bf B303}, 407-454.}
\item{}{Guilliani, D.  (1994) article in this volume.}
\item{}{Haji\v cek, P.  (1994) Article in this volume.}
\item{}{Heisenberg, W.  (1963) Interview by T.S. Kuhn of February 25th.}
\item{}{Hosoya, A. \& Nakao, K.  (1990) Class. Quan. Grav. {\bf 7} 163-176.}
\item{}{Iwasaki, J. \& Rovelli, C.  (1993) Int. J. Mod. Phys. {\bf D1}
533-558.}
\item{}{Kucha\v r  (1970) J. Math. Phys. {\bf 11} 3322-3334.}
\item{}{Loll, R.  (1992) Nucl. Phys. {\bf B368}, 121-142.}
\item{}{Loll, R.  (1994) article in this volume.}
\item{}{Marolf, D. (1993) Loop representations of 2+1 gravity on a torus,
SUGP-93/3/1.}
\item{}{Marolf D. \& Mour\~ao J.  (1993) Carrier space of the
Ashtekar-Lewandowski measure, in preparation.}
\item{}{Mena-Marug\'an, G. A.  (1993) Reality conditions for Lorentzian
and Euclidean gravity in the Ashtekar formulation, CGPG-93/9-3; Int.
J.  Mod. Phys. {\bf D} (in press).}
\item{}{Matschull, H.J. \& Nicolai, H. (1993) DESY 93-073; Nucl Phys.
{\bf B} (in press)}
\item{}{Mitter, P.K. \& Viallet C.  (1981) Commun. Math. Phys. {\bf 79}
43-58.}
\item{}{Moncrief, V.  (1989) J. Math. Phys. {\bf 30} 2907-2914.}
\item{}{Morales-T\'ecotl, H.A. \& Rovelli, C. (1994) University of
Pittsburgh pre-print.
\item{}{Nelson, J. \& Regge, T.  (1989) {\bf A4}, 2021-2030.}
\item{}{Nicolai, H. \& Matschull, H.J. (1993) J. Geo. \& Phys.
{\bf 11} 15-62}.
\item{}{Pais, A. (1987) In: {\it Paul Adrian Maurice Dirac},
B. N. Kursunoglu \& E. P. Wigner (eds), Cambridge University Press,
Cambridge, page 96.}
\item{}{Pullin, J.  (1993) Knot theory and quantum gravity in loop
space: A premier, hep-th/9301028, to appear in the proceedings of the
Vth Mexican school on particles and fields, J.L. Lucio (ed), Word
Scientific, Singapore.}
\item{}{Rovelli, C. \& Smolin, L.  (1990) Nucl. Phys. {\bf B331}, 80-152.}
\item{}{Rovelli, C. \& Smolin, L.  (1993)  The physical Hamiltonian in
non-perturbative quantum gravity, CGPC-93/8-2.}
\item{}{Smolin, L.  (1993) In: {\it General relativity and gravitation 1992},
 R.J. Gleiser et al (eds), Institute of Physics, Bristol, pages 229-261.}
\item{}{Streater, R. \& Wightman, A.  (1964) {\it PCT, spin and statistics,
and all that}, Benjamin, New York.}
\item{}{Thiemann, T.  (1994) article in this volume.}
\item{}{Witten, E.  (1988) Nucl. Phys. {\bf B311}, 46-78.}
\item{}{Zegwaard, J.  (1993) Phys. Lett. {\bf B378} 217-220.}
\item{}{Zegwaard, J.  (1994) {\it The loop representation of canonical
quantum gravity and its interpretation} Ph.D. thesis, University of Utrecht.}
\bye